\documentclass[fleqn,10pt]{wlscirep}
\usepackage[utf8]{inputenc}
\usepackage[T1]{fontenc}
\usepackage{comment}
\usepackage{here}
\usepackage{csvsimple}
\usepackage[subrefformat=parens]{subcaption}

\title{Dynamic Grass Color Scale Display Technique Based on Grass Length for Green Landscape-Friendly Animation Display}

\author[1,*]{Kojiro Tanaka}
\author[1]{Yuichi Kato}
\author[1]{\color{black}Akito Mizuno\color{black}}
\author[2]{Masahiko Mikawa}
\author[2]{Makoto Fujisawa}
\affil[1]{Graduate School of Comprehensive Human Sciences, University of Tsukuba, Japan}
\affil[2]{Faculty of Library Infromation and Media Science, University of Tsukuba, Japan}

\affil[*]{Mail: tanaka.kojiro.sp@alumni.tsukuba.ac.jp}

\begin{abstract}
  Public displays, such as liquid crystal displays (LCDs), are often used in urban green spaces. However, these display devices often tarnish the green landscape of urban green spaces due to their artificial materials. We previously proposed a green landscape-friendly grass animation display that dynamically controls the grass color pixel by pixel. The grass color is changed by moving a green grass length in yellow grass, and the grass animation display runs simple animations using grayscale images. In our previous study, the color scale is subjectively mapped to the green grass length. However, this method fails to display the grass colors corresponding to the color scale based on objective evaluations. Herein, we introduce a dynamic grass color scale display technique based on the grass length. We develop a grass color scale setting procedure to map the grass length to the five-level color scale through image processing. In the outdoor experiment of the grass color scale setting procedure, the color scale corresponds to the green grass length based on a viewpoint. Finally, we demonstrate a grass animation display to show the animations with the color scale using experimental results.
\end{abstract}
\begin{document}

\flushbottom
\maketitle
%
%
\thispagestyle{empty}


\section*{Introduction}


Technologies of smart cities have been researched and developed for multiple fields \cite{SILVA2018697}, and one of the missions of smart cities is to provide people with more opportunities to access information and improve the convenience of their community. Recently, public displays usually work in outdoor public spaces and can contribute to creating an environment that provides useful information to people easily \cite{surveyPervasive}. Liquid crystal displays (LCDs) and projectors are often used as public displays because they can show stable high-resolution animations and can be installed outdoor easily. An animation function allows public displays to provide multiple information quickly to people. In addition, people can interact with public displays using the animation function. In recent years, the effect of interactions between people and public displays has attracted much more attention, and intuitive interaction design in public spaces was discussed \cite{Strategies}. Numerous researchers and developers have exploblack the use of public displays in multiple scenarios, such as transportation \cite{alt2016Opportunistic, coenen2016Synchronized}, shopping \cite{yoshino2020KI,muta2015Interactive}, public facilities \cite{yoo2020Putting,kukka2018UbiLibrary}, sports \cite{claes2016Bicycle, coenen2021Public}, and health \cite{chang2014Lunch, altmeyer2018Extending}. Public displays have also been used in urban green spaces. For example, the digital signage displays in the Manchester Science Park provide users with information such as a map and event schedules \cite{FreeStanding}. Moreover, an art projection mapping was created in Sosei river park using some projectors \cite{northerncrossSosei}. However, current public display devices can spoil green landscapes of urban green spaces, as most of the display devices resemble artificial materials \cite{surveyPervasive}. Numerous researchers revealed that green landscapes have a positive impact on human well-being and health \cite{maller2006Healthy,abraham2010Landscape,lee2017Experimental}. Thus, green landscapes of urban green spaces must be preserved to enhance comfort in people's lives. Therefore, urban green spaces require a living environment-friendly display method to preserve their green landscapes. 

\begin{figure}[h]
  \centering
  \includegraphics[width=\linewidth]{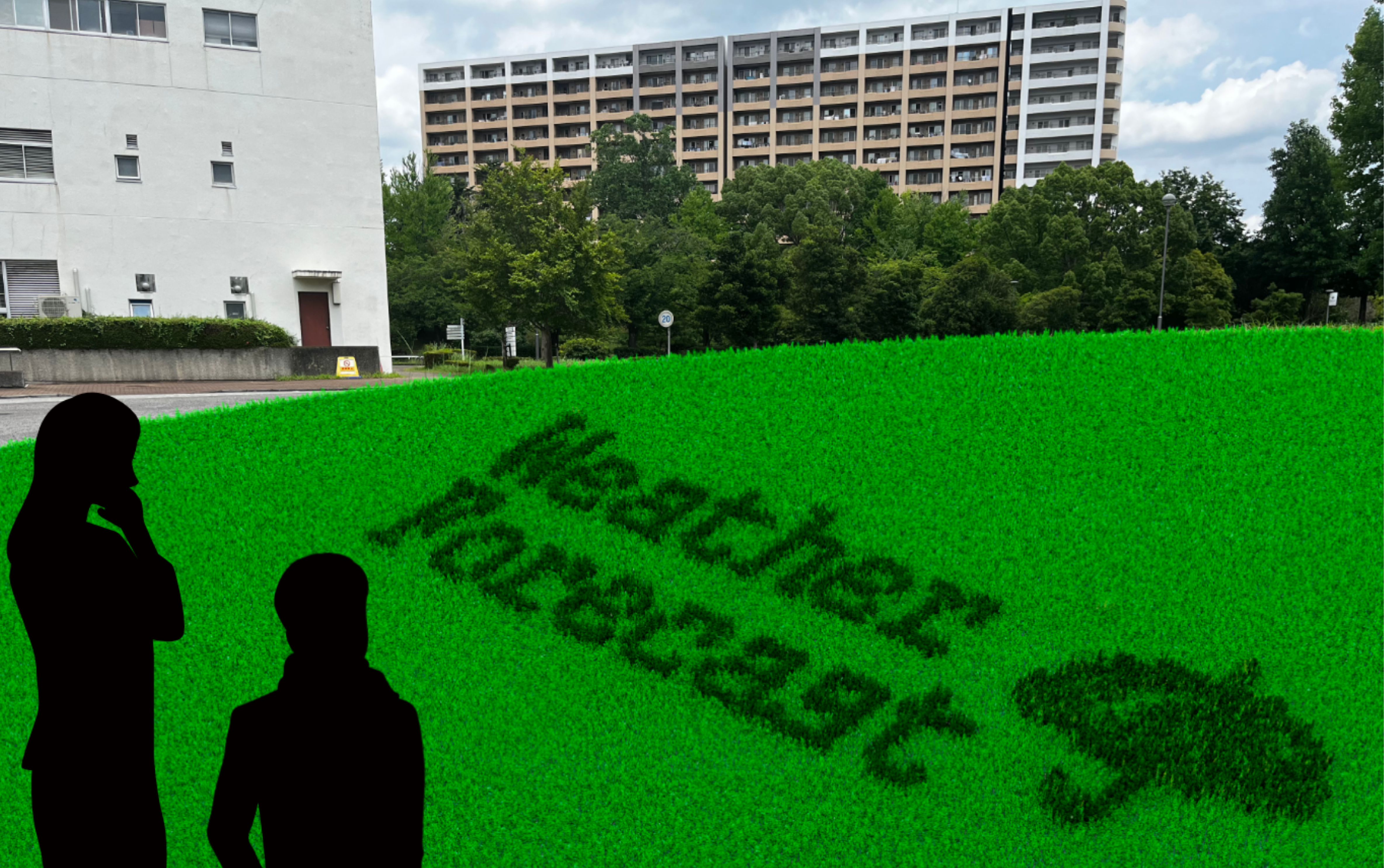}
  \caption{Concept of green landscape-friendly animation display in urban green space. The grass animation display runs a weather forecast application while preserving the green landscape of the urban green space. The background photo was captublack at University of Tsukuba. The umbrella material was used in SKY's animation (\url{https://pata2.jp}). This figure was created using Autodesk Maya version 2023 (\url{https://www.autodesk.com/products/maya/features}) and Serif Affinity Designer version 1.10.0 (\url{https://affinity.serif.com/en-us/}).}
  \label{application}
\end{figure}

Numerous researchers and artists have exploblack multiple living environment-friendly display methods using natural materials such as foodstuffs \cite{robinson2019Sustainabot}, wet ground \cite{nagafuchi2020Polka}, a fur carpet \cite{sugiura2014Graffiti}, and rice paddy \cite{2015Rice}. In particular, displaying images using grass has been studied as a green landscape-friendly display. All current grass display methods can show a static image. Sugiura et al. proposed a print device to draw a binary static image on a grass surface using anisotropic reflection \cite{sugiura2017Grassffiti}. The print device has multiple servo motor rods to raise or flatten the grass blades. Consequently, the grass surface can be brighter or darker pixel by pixel. There is a commercial grass service to print a static image on a large grass field, such as a soccer field \cite{AirPrint}. Because the direction of the grass blades can be adjusted by air pressure to change the grass brightness, the large static image can be displayed on the grass surface using a vehicle with an air pressure system. Scheible et al. developed a drone-aided system to create a large grass artwork. \cite{scheible2016DroneLandArt}. An artist can mow the grass to generate the artwork through a bird's eye view provided by several drones. In art fields, Ackroyd \& Harvey created an artwork that shows photography on the grass surface \cite{Big}. The grass surface is exposed to light projected from a negative image of the photography throughout the grass growth. Consequently, as the grass grows, the grass color is generated to show the photography. These grass methods can show the static images in the urban green spaces to preserve the green landscapes. However, these grass methods cannot display animations, as it is difficult to dynamically change the grass color. A shape-changing-grass system is requiblack to solve this problem. There are several smart material interfaces to bend artificial grass blades automatically actuated by shape memory alloys \cite{minuto2012Growing,ohkubo2013Interface}. However, these methods do not focus on displaying animations on the grass surface. Therefore, we focus on the smart city technology of a grass animation display system to show animations while it preserves the green landscape of the urban green spaces. For example, a landscape-friendly weather forecast application can be realized by applying the grass animation display as shown in Figure \ref{application}.

We previously proposed a green landscape-friendly display method that realizes animations on a grass surface by dynamically controlling grass length \cite{tanaka2021Natural}. This method focused on a grass color consisting dying yellow grass and lively green grass in the real world. The length of the yellow grass is fixed, whereas, the green grass length increases with time. Thus, multiple grass colors are generated depending on the green grass length. We designed a grass system to move artificial green grass length in artificial yellow grass using a linear actuator system pixel by pixel. The grass system was named a grass pixel. Thus, a grass animation display can be constructed by using multiple grass pixels, and the animations can be directly displayed on the grass surface without a displaying device, such as an LCD or a projector. In previous research, we developed the grass pixel with artificial yellow and green grass using a linear actuator system to control the green grass length. We also conducted a simple experiment to evaluate the grass color change of the grass pixel through image processing. Consequently, we confirmed that the grass color of the grass pixel can simply be changed from yellow to green while increasing the green grass length. Moreover, a $3\times3$ pixels grass animation display was created using nine grass pixels and demonstrated in several simple animations. $3\times3$ grayscale images were used to play the animations on the grass animation display. Because the grass color can be changed based on the green grass length, the color scale of the grayscale images was subjectively mapped to the green grass length in the demonstrations. However, we did not realize a grass color scale setting procedure to map the color scale to the green grass length based on objective evaluations.

\begin{figure}[h]
  \centering
  \includegraphics[width=\linewidth]{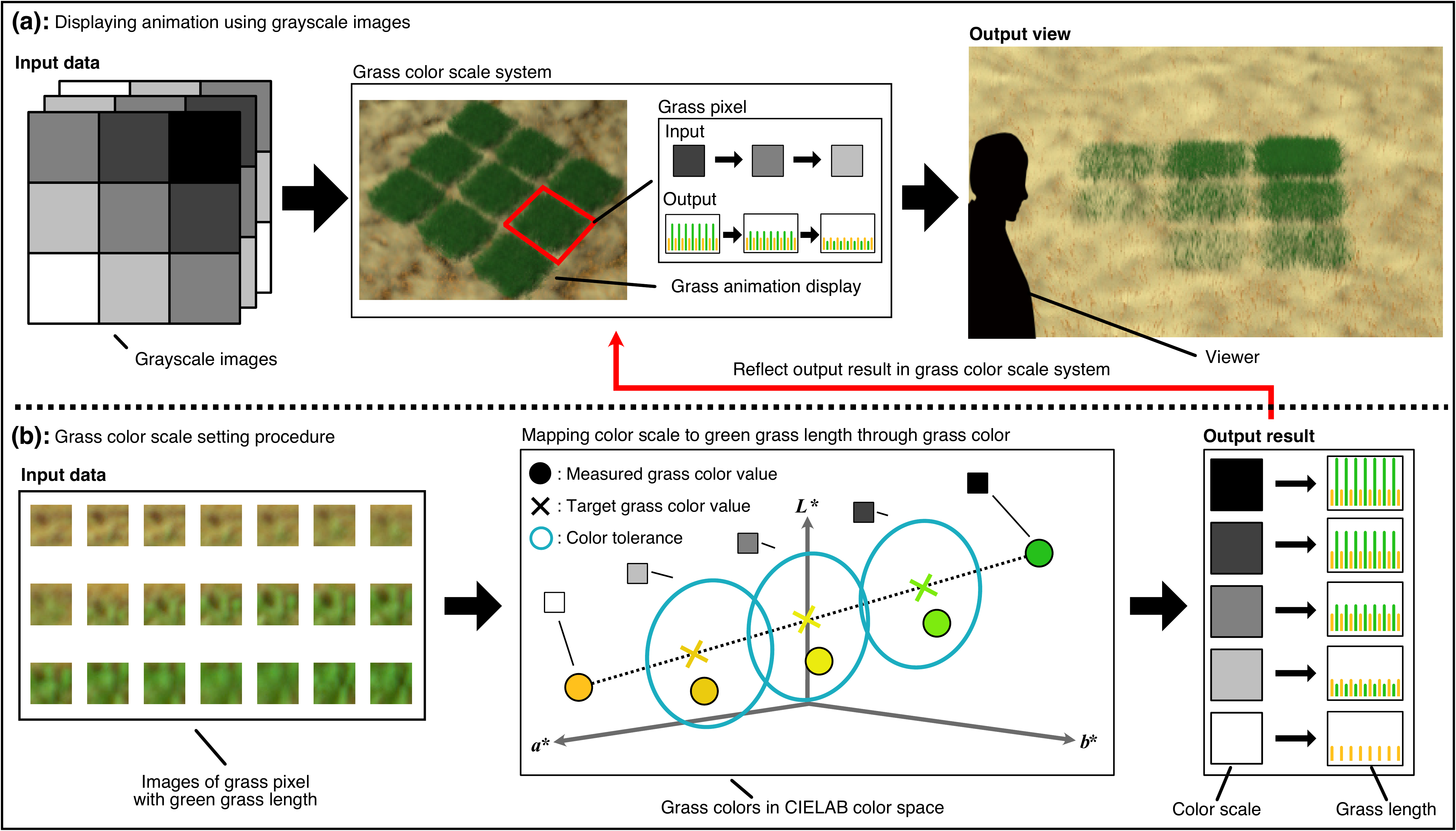}
  \caption{Concept of displaying grass color corresponding to color scale. (a) A grass color scale system allows a grass animation display to show an animation using grayscale images. The color scale corresponds to a green grass length. (b) Green grass length of the grass pixel corresponding to the color scale is determined using a grass color scale setting procedure. The results are reflected in the grass color scale system to show the animation on the grass animation display. This figure was created using Autodesk 3ds Max version 2021 (\url{https://www.autodesk.com/products/3ds-max/features}) and Serif Affinity Designer version 1.10.0 (\url{https://affinity.serif.com/en-us/})}
  \label{concept2}
\end{figure}

In this paper, we introduce a dynamic grass color scale display technique based on grass length for a green landscape-friendly animation display. Figure \ref{concept2} illustrates the concept of our proposed method. We focus on a grass color scale system that allows the animations to be played on the grass animation display, as shown in Figure \ref{concept2}(a). In the grass color scale system, the color scale of the input grayscale images is mapped to the green grass length. The grass colors based on the green grass length corresponding to the color scale must be evenly separated to ensure clear animations. We develop a grass color scale setting procedure to map the grass length to the color scale through image processing, as shown in Figure \ref{concept2}(b). The grass color scale setting procedure provides the green grass length for each level of the color scale using the grass colors of the grass pixel measublack by a digital camera at regular green grass length intervals. In the setting procedure, the CIE 1976 $L^*a^*b^*$ (CIELAB) color space \color{black} and the color difference formula of CIEDE2000 are \color{black} adopted to evaluate the grass color. To avoid an effect of a unique camera maker's image processing engine, the grass color is captublack as a raw image and measublack in the CIELAB color space through the sRGB color space. The measublack value is referblack to as a measublack grass color value. The measublack grass colors suitable for the color scale are obtained by comparing the measublack grass color values with target grass color values, which are CIELAB values of reference grass colors corresponding to the color scale. The target grass color values corresponding to minimum and maximum levels of the color scale are the measublack grass color values when the green grass lengths are minimum and maximum, respectively. Then, other target color values are calculated as CIELAB values that equally divide the \color{black} color difference of the CIEDE2000 \color{black} between the target grass color values corresponding to the minimum and maximum levels of the color scale. Further, each target grass color value has a color tolerance. When each color tolerance includes the measublack grass color value closest to each target grass color value, the measublack grass pixel is consideblack to be capable of displaying the grass colors corresponding to the color scale. Therefore, the color scale can be mapped to the green grass length using the measublack grass colors corresponding to the color scale. Then, the grass animation display can show the animation clearly by reflecting the results of the grass color scale setting procedure in the grass color scale system.

To achieve a simple grass color scale for the grass animation display, we focus on the grass pixel that show the grass color corresponding to a color scale of five levels with reference to the color scale of retro games. We conducted experiments with the grass color scale setting procedure in natural sunlight from 10:00 a.m. to 2:00 p.m., developing a grass pixel for our outdoor experiments. The green grass length was changed from 0.00 to 15.00 [mm], and the grass color of the grass pixel was measublack at regular intervals of 0.75 [mm]. Moreover, the experiments were conducted at multiple camera positions focusing on a viewer's height, a distance, and a horizontal angle between the viewer and the grass pixel. Through the experiments, the grass pixel possibilities of the color scale are clarified, such that the color scale corresponds to the green grass length for each camera position. \color{black} In addition, we conducted experiments focusing on the repeatability and reliability of the grass pixel. In the experiments, we revealed that the grass pixel could show the same grass color to several camera positions when the same grass length was repeatedly inputted in natural sunlight from 10:00 a.m. to 2:00 p.m. \color{black} After the experiments, we demonstrated several animations using the results of our grass color scale experiments. In the demonstrations, a $3\times3$ pixels grass animation display is built using nine grass pixels, then, four animations are shown on the animation grass display using $3\times3$ pixels grayscale images with the color scale of five levels.

\section*{Results and discussion}

\subsection*{Grass pixel design}

\begin{figure}[h]
  \centering
  \includegraphics[width=0.9\linewidth]{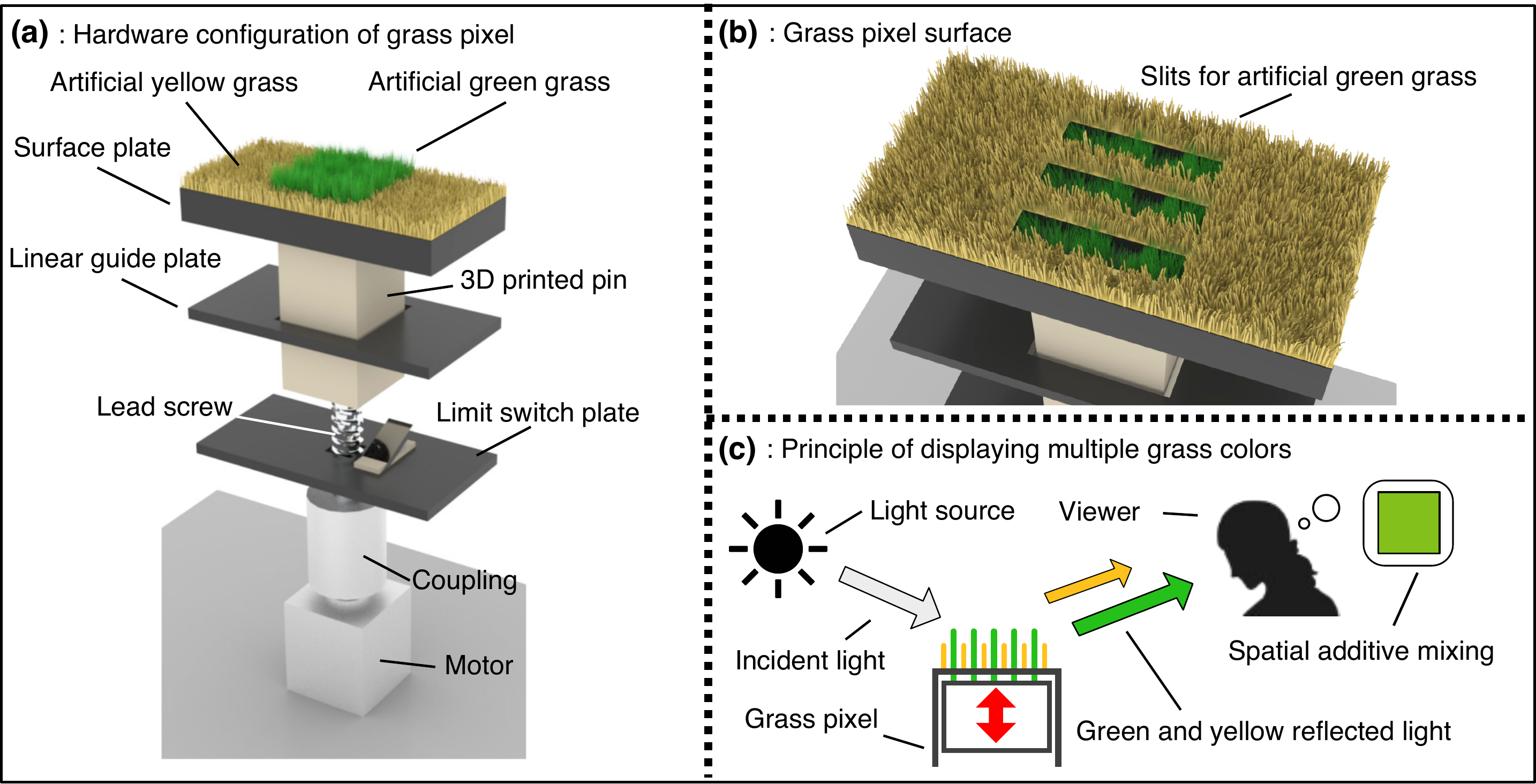}
  \caption{Design of grass pixel system. (a) A grass pixel is developed using a linear actuator system operated by a motor. Artificial green and yellow grass is planted on a top surface of a 3D printed pin and a surface plate, respectively. (b) The surface plate has slits to move the artificial green grass in and out of the artificial yellow grass. (c) The grass pixel displays multiple grass colors based on an area ratio of the artificial green and yellow grass through spatial additive mixing. This figure was created using Autodesk 3ds Max version 2021 (\url{https://www.autodesk.com/products/3ds-max/features}) and Serif Affinity Designer version 1.10.0 (\url{https://affinity.serif.com/en-us/})}
  \label{hardware}
  \vspace{-10pt}
\end{figure}

Figure \ref{hardware} shows an overview of a grass pixel design. The grass pixel system is developed based on a linear actuator system and can display multiple grass colors by dynamically moving the green grass length. As shown in Figure \ref{hardware}(a), the grass pixel consists of artificial green grass, artificial yellow grass, a surface plate, a linear guide plate, a limit switch plate, a three-dimensional (3D) printed pin, a coupling and, a motor. The motor is connected to the lead screw using the coupling, and the lead screw is jointed to the 3D printed pin. Then, the 3D printed pin with the linear guide plate can be moved vertically while the motor rotates the lead screw. The limit switch plate has a limit switch to set the home position of the 3D printed pin. The green grass is planted on the top surface of the 3D printed pin. As shown in Figure \ref{hardware}(b), the yellow grass is planted on the surface plate that has slits to move the green grass in and out the yellow grass as if lively green grass grows between the gaps of the dying yellow grass. Therefore, the green grass length can be changed by moving the 3D printed pin. The detailed grass pixel hardware for our experiment is described in the next subsection of the grass color scale experiments.

As shown in Figure \ref{hardware}(c), the grass pixel shows multiple grass colors according to the green grass length through spatial additive mixing, which is a type of additive color mixing. When light is incident on the grass pixel, green and yellow reflected lights based on the grass color are sent to the viewer. Then, the viewer recognizes an intermediate color between green and yellow when the viewer is at a distance too far away from the grass pixel to distinguish between the yellow and green grass. Moreover, the quantities of the green and yellow reflected lights can be changed according to the area ratio of the green to yellow grass. The area ratio can be changed based on the green grass length, therefore, the viewer observes multiple grass colors of the grass pixel through spatial additive mixing.

\subsection*{Grass color scale setting procedure of grass pixel}

\begin{figure}[h]
  \centering
  \includegraphics[width=0.72\linewidth]{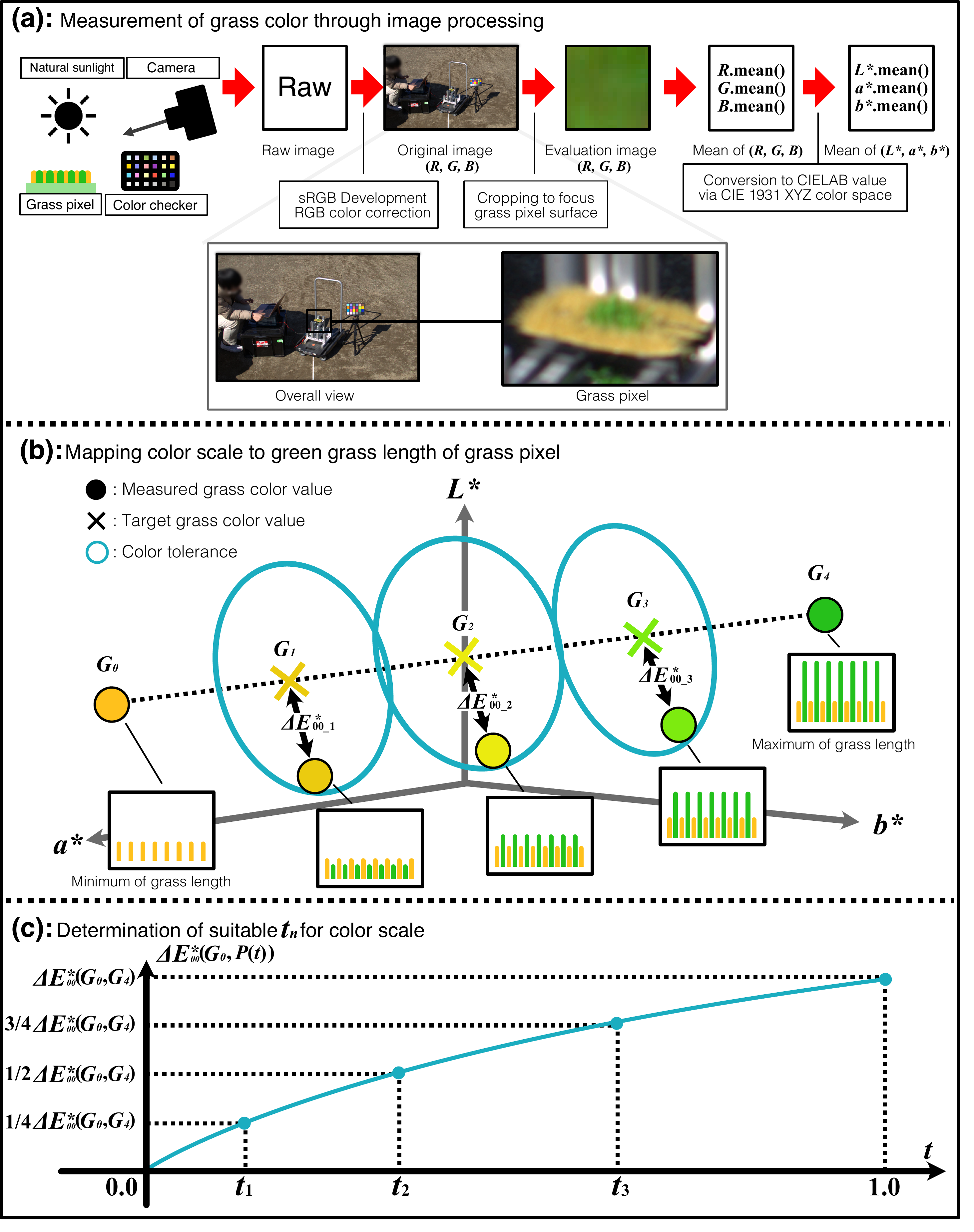}
  \vspace{-10pt}
  \caption{Overview of grass color scale setting procedure. (a) The grass color of a grass pixel is measublack as a CIELAB value through image processing. (b) Measublack grass color values are compablack with target grass color values to map the grass length of the grass pixel to a color scale with five levels. \color{black} (c) To calculate the target grass color value, $t_n$ suitable for $G_n$ is determined so that $\Delta E^*_{00} (G_0, P(t_n))$ is $(n/4) \cdot \Delta E^*_{00} (G_0, G_4)$. \color{black} This figure was created using Serif Affinity Designer version 1.10.0 (\url{https://affinity.serif.com/en-us/}).}
  \label{Define}
\end{figure}

Figure \ref{Define} illustrates an overview of the grass color scale setting procedure of the grass pixel. Herein, a color scale with five levels is mapped to the green grass length of the grass pixel through the grass color scale setting procedure. In the setting procedure, the grass colors of the grass pixel are measublack in the CIELAB color space, which is defined by International Commission on Illumination (CIE). $L^*$, $a^*$, and $b^*$ represent lightness, complementary colors of black ($+a*$) and green ($-a*$), and complementary colors of yellow ($+b^*$) and blue ($-b^*$), respectively. When a color is compablack with another color, the color difference usually represents a Euclidean distance. However, many color spaces, including HSV (Hue, Saturation, and Value of Brightness) color space, have the problem of differences in the perceived sense depending on the location of the color, even at the same color difference. In other words, the color difference that people perceive as different is expressed at different distances depending on the location of the color. In contrast, the CIELAB color space \color{black} had been \color{black} designed so that the same color difference provides the same perceptual difference at any color position. \color{black} A color difference using a Euclidean distance between two CIELAB values $\Delta E^*_{ab}$ had been defined as CIE76 in 1976 and is displayed below. \color{black}

\color{black}
\begin{eqnarray}
  \Delta E^*_{ab} = \sqrt{(L^*_2 - L^*_1)^2 + (a^*_2 - a^*_1)^2 + (b^*_2 - b^*_t)^2}
\end{eqnarray}

\noindent where $(L^*_1, a^*_1, b^*_1)$ and $(L^*_2, a^*_2, b^*_2)$ are example CIELAB values.

However, it was found that there was a difference between $\Delta E^*_{ab}$ and visual evaluation by humans, especially at the location of color with large saturation. Thus, in 2001, CIE published an alternative color difference formula called CIEDE2000 $\Delta E^*_{00}$ according to human perception. $\Delta E^*_{00} $ can be calculated based on lightness difference, saturation difference, and hue difference, and can be corrected with weighing coefficients ($S_L, S_C$, and $S_H$) and constants called parametric coefficients ($k_L, k_C$, and $k_H$). Sharma et al. presented the color difference of $\Delta E^*_{00}$ between two example CIELAB values $(L^*_1, a^*_1, b^*_1)$ and $(L^*_2, a^*_2, b^*_2)$ as below \cite{sharma2005ciede2000}.

\begin{eqnarray}
  \Delta E^*_{00} = \sqrt{
    \biggl(\dfrac{\Delta L'}{k_L S_L}\biggr)^2
  + \biggl(\dfrac{\Delta C'}{k_C S_C}\biggr)^2
  + \biggl(\dfrac{\Delta H'}{k_H S_H}\biggr)^2
  + R_T \dfrac{\Delta C'}{k_C S_C} \dfrac{\Delta H'}{k_H S_H}
  }
\end{eqnarray}

\noindent where $k_L, k_C$, and $k_H$ are determined depending on measurement conditions, however, all of the parameters are set to 1 on the standard conditions specified for CIEDE2000. $\Delta L', \Delta C'$, and $\Delta H'$ can be calculated using the differences of lightness, saturation, and hue, respectively as below.

\begin{eqnarray}
  \Delta L' &=& L^*_2 - L^*_1 \\ 
  \Delta C'  &=&C'_2 - C'_1 \\
  \Delta H' &=& 2 \sqrt{C'_1 C'_2} \sin (\dfrac{\Delta h'}{2}) \\
\end{eqnarray}

\noindent where

\begin{eqnarray}
  C'_i &=& \sqrt{a'^2_i + b^{*2}_i} \quad (i = 1, 2) \\
  a'_i  &=&(1+G)a^*_i \quad (i = 1, 2) \\
  G &=& 0.5 \Biggl(1 - \sqrt{\dfrac{\bar{C^*_{ab}}^7}{\bar{C^*_{ab}}^7+25^7}}\Biggr) \\
  \bar{C^*_{ab}} &=& \dfrac{C^*_{1,ab}+C^*_{2,ab}}{2}  \\
  C^*_{i,ab} &=& \sqrt{(a^*_i)^2+(b^*_i)^2} \quad (i = 1, 2) \\
  \Delta h' &=& \left\{ 
    \begin{array}{ll}
      0 & (C'_1 C'_2 = 0) \\
      h'_2 - h'_1 & (C'_1 C'_2 \neq 0 \quad |h'_2 - h'_1| \leq 180^{\circ}) \\
      (h'_2 - h'_1) - 360^{\circ} & (C'_1 C'_2 \neq 0 \quad (h'_2 - h'_1) > 180^{\circ}) \\
      (h'_2 - h'_1) + 360^{\circ} & (C'_1 C'_2 \neq 0 \quad (h'_2 - h'_1) < -180^{\circ}) 
    \end{array}
  \right. \\
  h'_i &=& \left\{
    \begin{array}{ll}
      0 & (b^*_i=a'_i=0) \\
      \tan^{-1}(b^*_i, a'_i) & (\text{otherwise})
    \end{array}
  \right. \quad (i = 1, 2)
\end{eqnarray}

\noindent In addition, the weighing coefficients ($S_L, S_C$, and $S_H$) and the rotation factor $R_T$ are determined as below.

\begin{eqnarray}
  S_L &=& 1 + \dfrac{0.015(\bar{L'} - 50)^2}{\sqrt{20+(\bar{L'}-50)^2}} \\
  S_C &=& 1 + 0.045 \bar{C'} \\
  S_H &=& 1 + 0.015 \bar{C'} T \\
  R_T &=& - \sin (2 \Delta \theta) R_C
\end{eqnarray}

\noindent where
\begin{eqnarray}
  \bar{L'} &=& (L^*_1+L^*_2) / 2 \\
  \bar{C'} &=& (C^*_1+C^*_2) / 2 \\
  \bar{h'} &=& \left\{
    \begin{array}{ll}
      \dfrac{h'_1+h'_2}{2} & (|h'_1-h'_2| \leq 180^{\circ} \quad C'_1 C'_2 \neq 0) \\
      \dfrac{h'_1+h'_2+360^{\circ}}{2} & (|h'_1-h'_2| > 180^{\circ} \quad (h'_1+h'_2) < 360^{\circ} \quad C'_1 C'_2 \neq 0) \\
      \dfrac{h'_1+h'_2-360^{\circ}}{2} & (|h'_1-h'_2| > 180^{\circ} \quad (h'_1+h'_2) \geqq 360^{\circ} \quad C'_1 C'_2 \neq 0) \\
      (h'_1+h'_2) & (C'_1 C'_2 = 0)
    \end{array}
  \right. \\
  T &=& 1 - 0.17 \cos (\bar{h'}-30^{\circ}) + 0.24 \cos (2\bar{h'}) + 0.32 \cos (3\bar{h'} + 6^{\circ}) - 0.20 \cos (4\bar{h'} - 63^{\circ}) \\
  \Delta \theta &=& 30 \exp\biggl\{-\biggl[\dfrac{\bar{h'}-275^{\circ}}{25}\biggr]^2\biggr\} \\
  R_C &=& 2 \sqrt{\dfrac{\bar{C'}^7}{\bar{C'}^7 + 25^7}}
\end{eqnarray}

\noindent In this paper, we adopted the color difference of CIEDE2000 to compare two CIELAB values of the grass colors and described the color difference as below.

\begin{eqnarray}
  \Delta E^*_{00} ((L^*_1, a^*_1, b^*_1), (L^*_2, a^*_2, b^*_2))
\end{eqnarray}

\noindent where $(L^*_1, a^*_1, b^*_1)$ and $(L^*_2, a^*_2, b^*_2)$ are two example CIELAB values of the grass colors. In addition, $k_L, k_C$, and $k_H$ were set to 1 because the grass color of the grass pixel was measublack under natural sunlight between 10:00 a.m. and 2:00 p.m.

\color{black}

Figure \ref{Define}(a) shows a measurement of the grass color of the grass pixel through image processing. The measurement is conducted outdoors between 10:00 a.m. to 2:00 p.m., because Sato et al. described natural sunlight at this period as suitable for evaluating colors \cite{sato1997Requirements}. A digital camera is used to capture the grass color in the setting procedure. To compare the grass colors taken by the digital camera, it is necessary to capture a camera device-independent image and correct the RGB value of the recorded image for \color{black} changes in a color temperature of natural sunlight \color{black}. First, the digital camera captures a raw image of the grass pixel to avoid the effect of the camera maker's unique image processing engine. Then, the raw image is developed in the sRGB (standard-RGB) color space, which is a camera device-independent RGB color space. Second, the RGB color correction of the developed image is performed using a color checker, which is a color calibration material consisting of multiple painted samples. When the digital camera captures the grass pixel, the color checker installed near the grass pixel is also captublack. Then, after developing the raw image, the RGB value of the developed image is corrected toward a reference image through the color checker. In the grass color scale setting procedure, the developed image when the green grass length is minimum is set as the color correction reference image. \color{black} In other words, the RGB correction allows the color temperatures of the other developed images to be corrected to the color temperature when the green grass length is minimum for each camera position. \color{black} Finally, an original image is created from the raw image through the sRGB development and the RGB color correction. The original image is independent of the camera device and takes into account natural sunlight changes. Because a calculation area focusing on the grass pixel surface must be decided to measure the grass color, the original image is cropped to create a square evaluation image, which contains only the part of the grass pixel surface. The evaluation image is used to calculate the grass color value in the CIELAB color space. The average RGB value of the evaluation image is calculated and converted to the CIELAB value via the CIE 1931 XYZ color space. Therefore, the grass color of the grass pixel can be measublack in the CIELAB color space, and the CIELAB value of the grass color is referblack to as the measublack grass color value. 

The grass colors \color{black} of the grass pixel \color{black} are measublack at regular green grass length intervals from minimum to maximum through image processing. Subsequently, a reference grass color value corresponding to the color scale is calculated using the measublack grass color values. \color{black} In this paper, we defined the reference grass color value as a target grass color value $G_n(L^*_n, a^*_n, b^*_n)$. In this case, $n$ represents a level of the color scale $(n =0, 1, 2, 3, 4)$. As shown in Figure \ref{Define}(b), both ends of the target grass color values $G_0$, $G_4$ denote the measublack grass color values when the green grass lengths are minimum and maximum, respectively. $G_1, G_2$, and $G_3$ are obtained from the CIELAB values on the line segment between $G_0$ and $G_4$. Then, let $P(t)$ be the CIELAB value on the line segment between $G_0$ and $G_4$ as below. 

\begin{eqnarray}
  P(t) = (L^*(t), a^*(t), b^*(t)) \quad (0 \leq t \leq 1)
\end{eqnarray}

\noindent where

\begin{eqnarray}
  L^*(t) &=& t(L^*_4 - L^*_0) + L^*_0  \\
  a^*(t) &=& t(a^*_4 - a^*_0) + a^*_0 \quad (0 \leq t \leq 1)\\
  b^*(t) &=& t(b^*_4 - b^*_0) + b^*_0  
\end{eqnarray}

\noindent $t$ represents a variable for which $P(t)$ is the CIELAB value on the line segment between $G_0$ and $G_4$ and has a definition range from 0 to 1. $(L^*_0, a^*_0, b^*_0)$ and $(L^*_4, a^*_4, b^*_4)$ are the target grass color values $G_0$ and $G_4$, respectively. To obtain the suitable $G_n$, it is needed to calculate $t$ to maintain color differences between neighboring $G_n$ constant. 

Figure \ref{Define}(c) shows how to determine $t$ for $G_n$. This graph of Figure \ref{Define}(c) represents the function of the color difference $\Delta E^*_{00}(G_0, P(t))$ with respect to $t$. Since $P(t)$ gets closer to $G_4$ from $G_0$ when $t$ increases, $\Delta E^*_{00}(G_0, P(t))$ rises from 0 to $\Delta E^*_{00}(G_0, G_4)$. $t$ suitable for $G_n$ can be uniquely obtained when $\Delta E^*_{00}(G_0, P(t))$ simply increases in the range of $0 \leq t \leq 1$. However, $\Delta E^*_{00}(G_0, P(t))$ is nonlinear with respect to $t$ because the color difference of CIEDE2000 is not linear with respect to a Euclidean distance in the CIELAB color space. Therefore, $\Delta E^*_{00}(G_0, P(t))$ is requiblack as below to clarify whether $\Delta E^*_{00}(G_0, P(t))$ can increase simply. When the condition below is satisfied, $t$ suitable for $G_n$ can be uniquely calculated.

\begin{eqnarray}
  \frac{d}{dt} \Delta E^*_{00}(G_0, P(t)) > 0 \quad (0 \leq t \leq 1)
  \label{partial}
\end{eqnarray}

\noindent Herein, let $t_n$ be the unique and suitable $t$ for $G_n$. To maintain the color differences between neighboring $G_n$ constant, $t_n$ and the CIELAB values of $G_n$ can be obtained as below. In addition, the determination of $t_n$ is also shown in Figure \ref{Define}(c).

\begin{eqnarray}
  \Delta E^*_{00}(G_0, P(t_n)) &=& \dfrac{n}{4} \Delta E^*_{00}(G_0, G_4) \quad (n = 0, 1, 2, 3, 4) \\
  G_n = P(t_n) &=& (L^*(t_n), a^*(t_n), b^*(t_n)) \quad (n = 0, 1, 2, 3, 4)
\end{eqnarray}

Figure \ref{Define}(b) shows how to map the color scale to the green grass length of the grass pixel. \color{black} $G_1, G_2$, and $G_3$ have a color tolerance that is a range of the certain color difference centeblack at each target grass color value as shown by blue \color{black}ellipses \color{black} in Figure \ref{Define}(b). When a measublack grass color value is within a color tolerance, the measublack and target grass colors are consideblack to be almost the same color. In the grass color scale setting procedure, \color{black} we adopted the category of the color tolerance created by ANSI and Committee for Graphic Arts Technologies Standards (CGATS) \cite{an2012graphic}. The color tolerance category is called CGATS TR 016-2014 and describes several levels of the color tolerance based on CIEDE2000. In the procedure, to compare the target grass color value with the measublack grass color value on an industrial-level color standard, the color tolerance of $G_1, G_2$ and $G_3$ is set to the color tolerance of 3.0 which focuses on color-critical applications, e.g., commercial printing. \color{black} In this case, the target grass color value must not be included in other color tolerances to distinguish the target grass colors, then, the color difference between $G_0$ and $G_4$ must be higher than \color{black}12\color{black}. To reveal whether the measublack grass color value is within the color tolerance, the color difference \color{black} $\Delta E^*_{00\_n}$ \color{black} between the measublack grass color value and $G_n$ is needed. The equation of \color{black} $\Delta E^*_{00\_n}$ \color{black} is displayed below.

\color{black}

\begin{eqnarray}
  \Delta E^*_{00\_n} = \Delta E^*_{00}(G_n, (L^*_m, a^*_m, b^*_m)) \quad (n = 1, 2, 3)
\end{eqnarray}

\noindent where $(L^*_m, a^*_m, b^*_m)$ is the measublack grass color value. Then, $\Delta E^*_{00\_1}, \Delta E^*_{00\_2}$, and $\Delta E^*_{00\_3}$ are calculated and compablack with the color tolerance of 3.0. When the $\Delta E^*_{00\_n}$ of the measublack grass color value is smaller than this color tolerance, the measublack grass color value can be almost the same color as $G_n$. Therefore, the grass pixel is regarded as displaying the grass colors corresponding to the color scale by adopting each grass length whose $\Delta E^*_{00\_n}$ is smaller than the color tolerance. In addition, the green grass length can likewise be mapped to the color scale. The detailed experimental environment of the grass color scale setting procedure is described in the next subsection of the grass color scale experiments.

\color{black}
\subsection*{Grass color scale experiments}

\subsubsection*{Experimental \color{black}e\color{black}nvironment}

\vspace{-5pt}

\begin{figure}[H]
  \centering
  \includegraphics[width=0.9\linewidth]{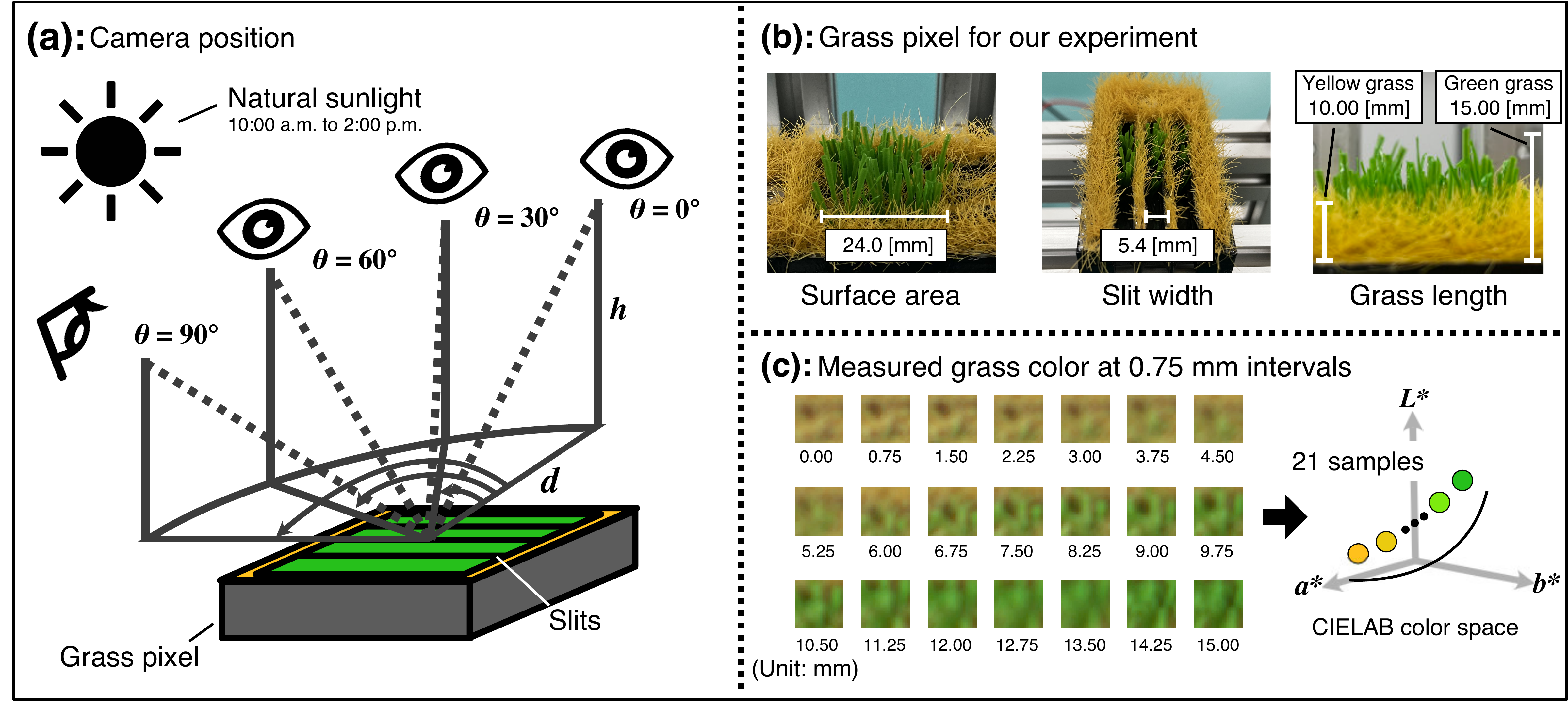}
  \caption{Experimental environment of grass color scale setting procedure. (a) A digital camera was located at multiple positions based on the viewer's height, a distance and a horizontal angle between a viewer and a grass pixel. (b) The grass pixel was developed for our experiment of the grass gradation scale setting procedure. (c) The grass color of the grass pixel was measublack as a CIELAB value at 0.75 mm intervals. This figure was created using Serif Affinity Designer version 1.10.0 (\url{https://affinity.serif.com/en-us/})}
  \label{experiment}
  \vspace{-10pt}
\end{figure}

We conducted an evaluation experiment to map the color scale to the green grass length using the grass color scale setting procedure. Figure \ref{experiment}(a) shows our experimental environment based on multiple camera positions. We used a digital camera, Nikon D7000 (resolution: 16.9 [MP]), with a single focus lens (AF-S DX NIKKOR 35 mm f/1.8G, focal length: 35 [mm], maximum aperture: f/1.8). The camera's aperture (f/4.0), shutter speed (1/640), and International Organization for Standardization speed (ISO100) were fixed. In addition, a color checker (Datacolor Spyder CHECKER 24) was installed near the grass pixel to correct the RGB value of the captublack image. The experiment was conducted on the playground of the University of Tsukuba in natural sunlight from 10:00 a.m. to 2:00 p.m. As shown in Figure \ref{experiment}(a), $h, d$, and $\theta$ represent the camera height from the grass pixel surface, the distance between the grass pixel and the digital camera, and the horizontal angle between them, respectively. Because the viewers are assumed to be an adult and a child, $h$ was set to 1.2, 1.6, and 1.7 [m]. $d$ was set to 1.0, 2.0, and 3.0 [m] as a distance to observe the grass color of the grass pixel.  $\theta$ was set to $0^\circ$, $30^\circ$, $60^\circ$, and $90^\circ$ around the center of the grass pixel. Furthermore, when $\theta$ was $0^\circ$ and $90^\circ$, the digital camera was perpendicular and parallel to the slits of the grass pixel, respectively. Figure \ref{experiment}(b) shows the grass pixel configuration in our grass color scale experiment. The surface area of the grass pixel was $24\times24$ [mm], and the slit width of the yellow grass surface was 5.4 [mm]. The artificial yellow grass length was 10.00 [mm], and the artificial green grass length could be varied from 0.00 to 15.00 [mm]. The grass pixel in our experiment consisted of a stepper motor (SM-42BYG011-25, 200 [step/rotation]), a lead screw with a coupling (length: 100.0 [mm], pitch: 6.0 [mm], lead: 6.0 [mm]), a 3D printed pin (dimension: $24\times24\times70$ [mm]), and a limit switch. Moreover, a limit switch plate, a guide plate, and a surface plate were created by a laser cutting device and a 3D printing device. The grass pixel was operated by a microcontroller (Teensy 3.5) via a stepper motor driver (DRV8825). The grass color of the grass pixel was measublack for each camera position while moving the grass length at 0.75 [mm] intervals from 0.00 to 15.00 [mm], as shown in Figure \ref{experiment}(c). Then, the color scale was mapped to the green grass length using the measublack grass color values for each camera position through the grass color scale setting procedure.

\vspace{-7pt}

\subsubsection*{Experimental \color{black}r\color{black}esults}
\begin{figure}[h]
  \centering
  \includegraphics[width=\linewidth]{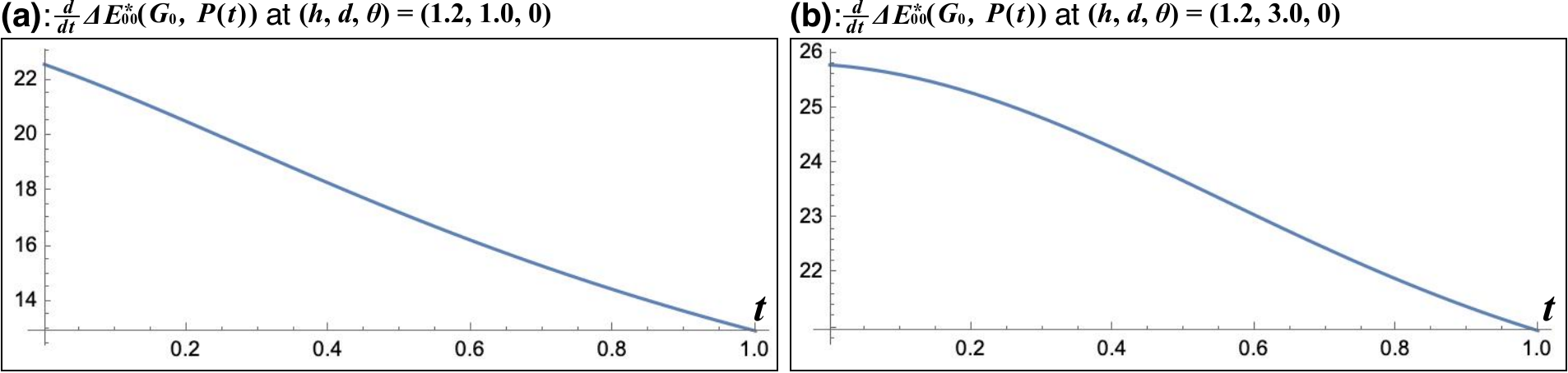}
  \caption{\color{black}Examples of $\tfrac{d}{dt}\Delta E^*_{00}(G_0, P(t))$. (a)(b) $\Delta E^*_{00}(G_0, P(t))$ can simply increase and $G_1, G_2$ and $G_3$ can be uniquely obtained at $(h, d, \theta) = (1.2, 1.0, 0)$ and $(1.2, 3.0, 0)$ because these resutls of $\tfrac{d}{dt}\Delta E^*_{00}(G_0, P(t))$ were positive in the range of $0 \leq t \leq 1$. This figure was created using Serif Affinity Designer version 1.10.0 (\url{https://affinity.serif.com/en-us/}) and Mathematica version 13.1.0.0 (\url{https://www.wolfram.com/mathematica/?source=nav}).\color{black}}
  \label{diff}
\end{figure}

\begin{figure}[H]
  \centering
  \includegraphics[width=0.87\linewidth]{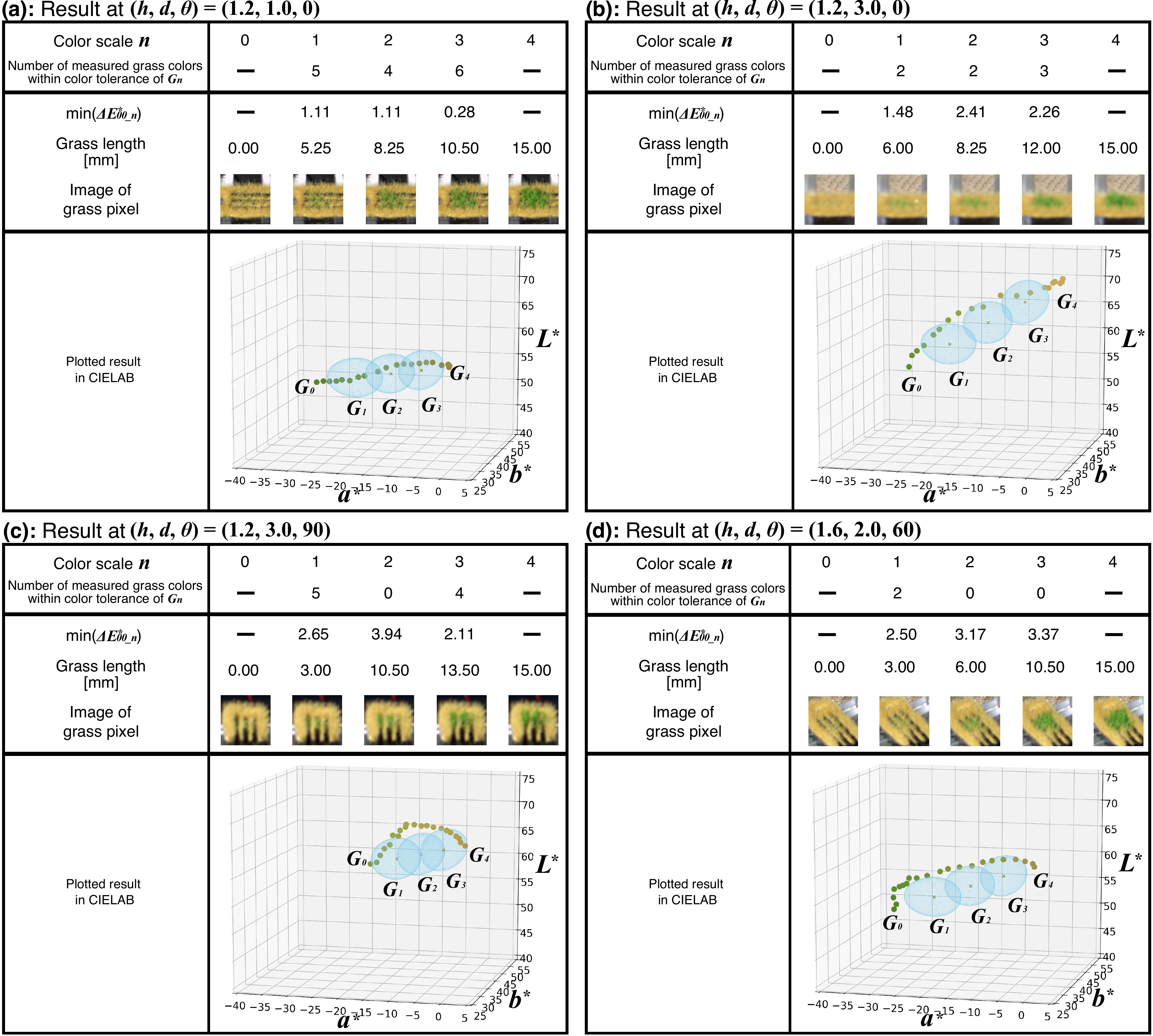}
  \caption{Examples of experimental results of grass color scale setting procedure. (a)(b) Results at $(h, d, \theta) = (1.2, 1.0, 0)$ and $(1.2, 3.0, 0)$ were successful to map the green grass length corresponding to the color scale. (c)(d) Results at \color{black} $(h, d, \theta) = (1.2, 3.0, 90)$ and $(1.6, 2.0, 60)$ \color{black} failed due to the influence of the slits of the grass pixel. This figure was created using Serif Affinity Designer version 1.10.0 (\url{https://affinity.serif.com/en-us/}) and Matplotlib version \color{black} 3.2.2 \color{black} (\url{https://matplotlib.org}).}
  \label{result}
  \vspace{-10pt}
\end{figure}

As a result, the grass pixel is regarded as displaying the grass colors corresponding to the color scale, and the green grass length was mapped to the color scale at \color{black} 30 \color{black} camera positions out of 36. \color{black} In these experiments, since $\tfrac{d}{dt}\Delta E^*_{00}(G_0, P(t))$ were positive in the range of $ 0 \leq t \leq 1$ in all camera positions, $G_1, G_2$ and $G_3$ can be uniquely obtained for each camera position, respectively. For example, Figure \ref{diff} shows $\tfrac{d}{dt} \Delta E^*_{00}(G_0, P(t))$ at $(h, d, \theta) = (1.2, 1.0, 0), (1.2, 3.0, 0)$. Since both results of $\tfrac{d}{dt}\Delta E^*_{00}(G_0, P(t))$ were positive even though these decreased in the range of $0 \leq t \leq 1$, $G_1, G_2$ and $G_3$ can be uniquely obtained at $(h, d, \theta) = (1.2, 1.0, 0), (1.2, 3.0, 0)$. Figure \ref{result} shows examples of the experimental results. These examples illustrate the number of measublack grass color values included within the color tolerance of $G_1, G_2$, and $G_3$ out of 21 measublack grass color values, respectively. In other words, the number represents the number of the grass length of the grass pixel corresponding to each level of the color scale. When this number is greater than or equal to 1 for each of $n=1, 2, 3$, the color scale with five levels can be mapped to the green grass length of the grass pixel. In addition, as examples of results of mapping the measublack grass colors of the grass pixel to the color scale, Figure \ref{result} also shows the color difference of $\Delta E^*_{00\_n}$, the grass lengths and the images of the grass pixel based on the measublack grass color values closest to $G_1, G_2$ and $G_3$, respectively. In this case, the color differences of $\Delta E^*_{00\_1}, \Delta E^*_{00\_2}$ and $\Delta E^*_{00\_3}$ are minimal. Furthermore, Figure \ref{result} shows plotted measublack and target grass color values in the CIELAB color space. Each dot marks the measublack grass color value and represents the measublack grass color. Each cross mark is the target grass color value and represents the target grass color. The blue \color{black} ellipsoid \color{black} centeblack on the cross mark represents the color tolerance of \color{black}3.0\color{black}. 

Figure \ref{result}(a) shows an example of the successful experimental result when $(h, d, \theta)$ was (1.2, 1.0, 0). In this result, the minima of \color{black}$\Delta E^*_{00\_1}, \Delta E^*_{00\_2}$ and $\Delta E^*_{00\_3}$\color{black} were within the color tolerances of $G_1, G_2$ and $G_3$, respectively. \color{black} Number of the measublack grass color values included in the color tolerance of $G_1, G_2$, and $G_3$ were 5, 4, and 6, respectively.  When the measublack grass color values based on the minima of $\Delta E^*_{00\_1}, \Delta E^*_{00\_2}$ and $\Delta E^*_{00\_3}$ are adopted, \color{black} the green grass lengths corresponding to the color scale can be 0.00, 5.25, \color{black} 8.25 \color{black}, \color{black} 10.50 \color{black}, and 15.00 [mm] in the order of the color scale. As shown in Figure \ref{result}(b), the result at $(h, d, \theta) = (1.2, 3.0, 0)$ was likewise successful. \color{black} Number of the measublack grass color values included in the color tolerance of $G_1, G_2$, and $G_3$ were 2, 2, and 3, respectively. In addition, the green grass lengths can be 0.00, 6.00, 8.25, 12.00, and 15.00 [mm] in the order of the color scale when the measublack grass color values based on the minima of $\Delta E^*_{00\_1}, \Delta E^*_{00\_2}$ and $\Delta E^*_{00\_3}$ are used. \color{black} Moreover, the plotted images show that some measublack grass color values are within the color tolerances \color{black} as shown in Figure \ref{result}(a) and (b). \color{black}

However, the slits of the grass pixel affected the experimental results at camera positions where they were captublack in the evaluation images. While the effects of the slits based on $h$ were rarely observed in the experiments, the results based on $d$ and $\theta$ were considerably influenced by \color{black} the slits \color{black}. The intensity of the measublack grass color values at $d=1.0$ was lower than at $d=2.0$ and 3.0 because the black slits were often captublack in the evaluation image when the digital camera and the grass pixel were at short distances to each other. For example, as shown in Figure \ref{result}(a) and (b), the dot colors at $(h, d, \theta) = (1.2, 1.0, 0)$ were darker than at $(h, d, \theta) = (1.2, 3.0, 0)$. \color{black} When $\theta$ were $60 ^\circ$ and \color{black} $90^\circ$, \color{black} the slits appeablack strongly in the evaluation image because the horizontal angle between the digital camera and the slits \color{black} was closer to \color{black} parallel \color{black} when $\theta $ increased. \color {black}Then, the trajectory of the measublack grass color values at \color{black} $\theta=60 ^\circ$ and \color{black} $90^\circ$ was more curved than at $\theta=0^\circ$ and $30^\circ$. For example, as shown in Figure \ref{result}(b) and (c), the plotted results at $(h, d, \theta) = (1.2, 3.0, 90)$ were more curved than at $(h, d, \theta) = (1.2, 3.0, 0)$. \color{black} Then, $G_2$ at $(h, d, \theta) = (1.2, 3.0, 90)$ cannot be mapped to the measublack grass color value because the minimum of $\Delta E^*_{00\_2}$ was 3.94, which were higher than the color tolerance of 3.0. In addition, the minima of $\Delta E^*_{00\_2}$ and $\Delta E^*_{00\_3}$ at $(h, d, \theta) = (1.6, 2.0, 60)$ were also 3.17 and 3.37 over the color tolerance of 3.0, respectively as shown in Figure \ref{result}(d). Furthermore, \color{black} the results at $(h, d, \theta) = (1.2, 2.0, 90)$, $(1.6, 3.0, 90)$ and $(1.7, 3.0, 60)$ likewise \color{black} failed to map the green grass length to \color{black} all the levels \color{black} of the color scale \color{black} simultaneously \color{black}. These results were attributed to the curved trajector\color{black}ies \color{black} of the measublack grass color values caused by \color{black} the \color{black} black slits in the evaluation images. \color{black} Moreover, $\Delta E^*_{00} (G_0, G_4)$ at most camera positions were over 12, however, $\Delta E^*_{00} (G_0, G_4)$ at $(h, d, \theta) = (1.6, 1.0, 90)$ was 11.94 under 12. Then, the measublack grass color values corresponding to the color scale cannot be determined because the target grass color value was included in the color tolerance of other target grass color values. The camera position of the result included $d=1.0$ and $\theta=90^\circ$, then, we suppose the result was caused by the color difference between $G_0$ and $G_4$ blackuced by the black slits of the grass pixel. \color{black}

\subsection*{Repeatability and reliability of grass pixel}

\begin{figure}[h]
  \centering
  \includegraphics[width=0.8\linewidth]{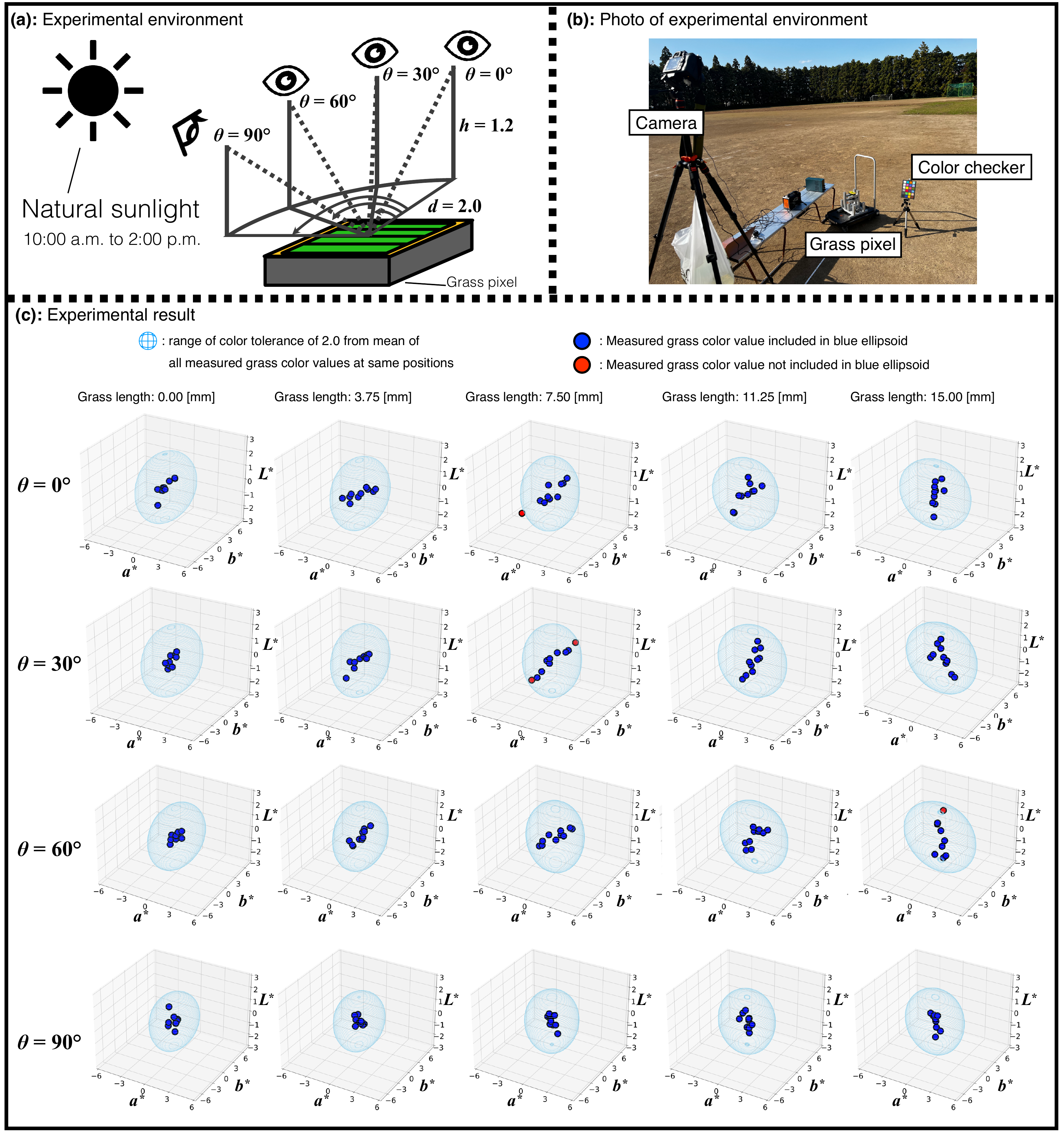}
  \caption{Simple experiments on repeatability and reliability of grass pixel. (a)(b) The camera height and the distance between the camera and the grass pixel were fixed. The horizontal angles between them were $0^\circ$, $30^\circ$, $60^\circ$, and $90^\circ$. \color{black} The experiments were conducted the playground in the University of Tsukuba under natural sunlight from 10:00 a.m. to 2:00 p.m. \color{black} (c) Since most of the measublack grass color values were within the blue \color{black} ellipsoid \color{black} of \color {black} the color tolerance of \color{black} the reference grass color value for each camera position, the experimental results revealed that the grass pixel can show the grass color stably with the grass length. This figure was created using Serif Affinity Designer version 1.10.0 (\url{https://affinity.serif.com/en-us/}) and Matplotlib version \color{black} 3.2.2 \color{black} (\url{https://matplotlib.org}). }
  \label{repeat}
\end{figure}

We conducted simple experiments on revealing whether the grass pixel can show the same grass color when the same grass length was repeatedly inputted to confirm the repeatability and reliability of the grass pixel. Figure \ref{repeat}(a) shows the experimental environment based on the grass color scale experiments \color{black} and Figure \ref{repeat}(b) shows the photo of the environment of the experiments. \color{black} In the experiments, the same digital camera and the grass pixel were used as in the grass color scale experiments. $h, d$, and $\theta$ represent the height of the digital camera, the distance from the digital camera to the grass pixel, and the horizontal angle between them as in the grass color scale experiment, respectively. $h$ and $d$ were 1.2 [m] and 2.0 [m]. Moreover, $\theta$ were $0^\circ$, $30^\circ$, $60^\circ$, and $90^\circ$. \color{black} The experiments were conducted on the playground of the University of Tsukuba in natural sunlight from 10:00 a.m. to 2:00 p.m. \color{black} The grass colors were measublack while moving the grass length in the following order: 0.00, 3.75, 7.50, 11.25, and 15.00 [mm]. Then, this process was repeated 10 times for each camera position. After the measurements, the mean of the measublack grass color values for each grass length and camera position was calculated as a reference grass color value. Then, the reference grass color value was compablack with 10 measublack grass color values at the same grass length and camera position. In the experiments, \color{black} to reveal the repeatability and reliability of the grass pixel using the tightest color tolerance, \color{black} the color tolerance of the reference grass color values was set to a color difference of \color{black} 2.0, which \color{black} is defined by \color{black} CGATS TR 016-2014 as a color difference for the most color-critical applications, e.g., proofing \cite{an2012graphic}. \color{black} In other words, when the measublack grass color values are more within the color tolerance, the grass colors of the measublack grass color values can be approximately the same and the grass pixel has repeatability and reliability.

Figure \ref{repeat}(c) shows the experimental results when $\theta$ were $0^\circ$, $30^\circ$, $60^\circ$, and $90^\circ$. In these graphs, the reference grass color value for each grass length and camera position is set as the origin. \color{black}T\color{black}he upper and the lower limits of \color{black} $a^*$ and $b^*$ are set to 6 and -6. In addition, them of $L^*$ is set to 3 and -3. \color{black} In other words, each parameter in these graphs represents the difference of each parameter between the reference and measublack grass color values. The blue \color{black} ellipsoid \color{black} of each graph is the color tolerance \color{black} of 2.0 \color{black} of the reference grass color value. Blue dots are the measublack grass color values within the color tolerance, and black dots are not included in the color tolerance. As shown in Figure \ref{repeat}(c), most of the measublack grass color values were included in the color tolerance of \color{black} 2.0 \color{black} at all angles. Although some of the measublack grass color values were not within the color tolerance, they were located \color{black} around the blue ellipsoid of the color tolerance. \color{black} Therefore, since the grass pixel can show the grass color stably with the grass length, the experiments revealed the repeatability and reliability of the grass pixel.

\subsection*{Demonstration for grass animation display}

\begin{figure}[h]
  \centering
  \includegraphics[width=\linewidth]{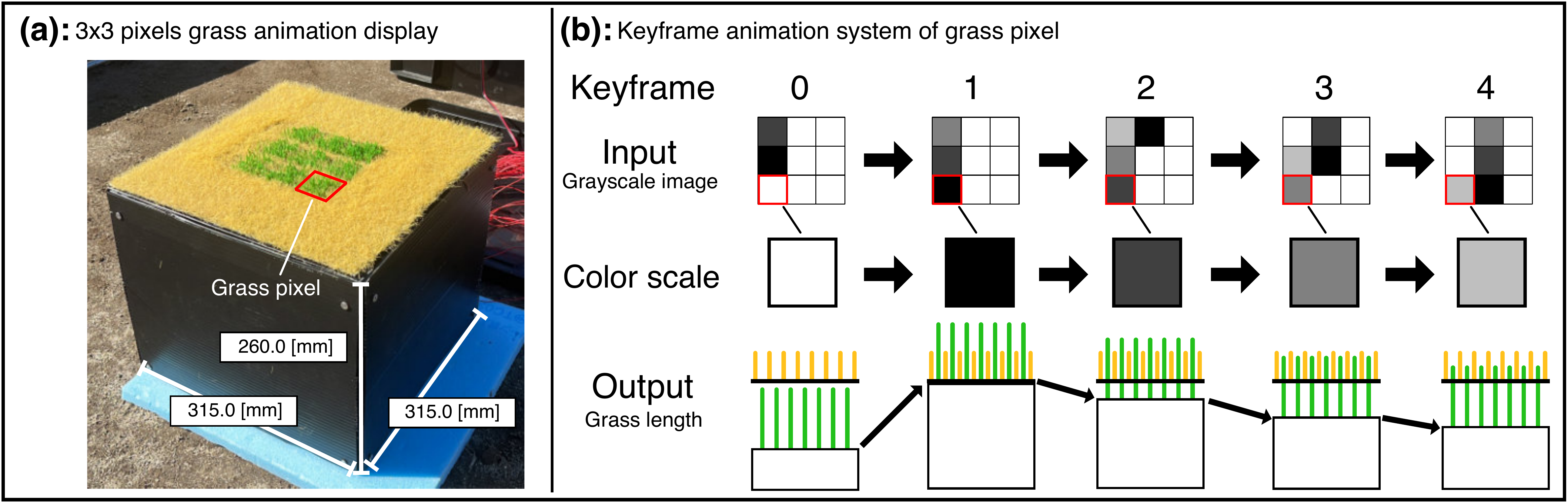}
  \caption{Animation system of a grass animation display used for demonstration. (a) A $3\times3$ pixels grass animation display was developed using nine grass pixels. (b) A keyframe animation system was adopted to play an animation on the grass animation display. A green length of a grass pixel was changed according to input levels of a color scale in keyframe order. This figure was created using Serif Affinity Designer version 1.10.0 (\url{https://affinity.serif.com/en-us/})}
  \label{animationsystem}
\end{figure}

We made several demonstrations of the grass animation display using the results of our grass color scale experiments. Figure \ref{animationsystem}(a) shows a $3\times3$ pixels grass animation display was constructed using nine grass pixels (dimension: $315.0\times315.0\times260.0$). In this display, the surface area of the grass pixel was $40\times40$ [mm]. The slit width and the moving range of the green grass length were the same as those of the grass pixel used in our grass color scale experiments. Because the color scale was mapped to the green grass length through the grass color scale experiments, the grass animation display can show a $3\times3$ pixels image using a $3\times3$ pixels grayscale image. Then, the animation was played on the grass animation display by changing the displayed images to other images with time. In the grass animation display, a keyframe animation system was adopted to change the displayed images as shown in Figure \ref{animationsystem}(b). In the animation system, the grayscale images were set in the order of the keyframes, and the levels of the color scale for each grass pixel were set in keyframe order. Then, the green grass length was determined pixel by pixel for each keyframe using the results of the grass color scale experiments. Therefore, each grass pixel can change the grass color corresponding to the color scale by moving its green grass length in keyframe order, and the grass animation display can play the animations by changing the displayed images. In the demonstrations, we captublack the grass animation display at $(h, d, \theta)=(1.7, 2.0, 0)$ and \color{black} adopted the green grass lengths suitable for the color scale. In the grass color scale experiments at the same camera position, the measublack grass color value of these green grass lengths was included within the color tolerance of 3.0 of each $G_n$. \color{black} The grass animation display played four animations, and the keyframe interval of the animation system was 1.0 [s].

\begin{figure}[h]
  \centering
  \includegraphics[width=\linewidth]{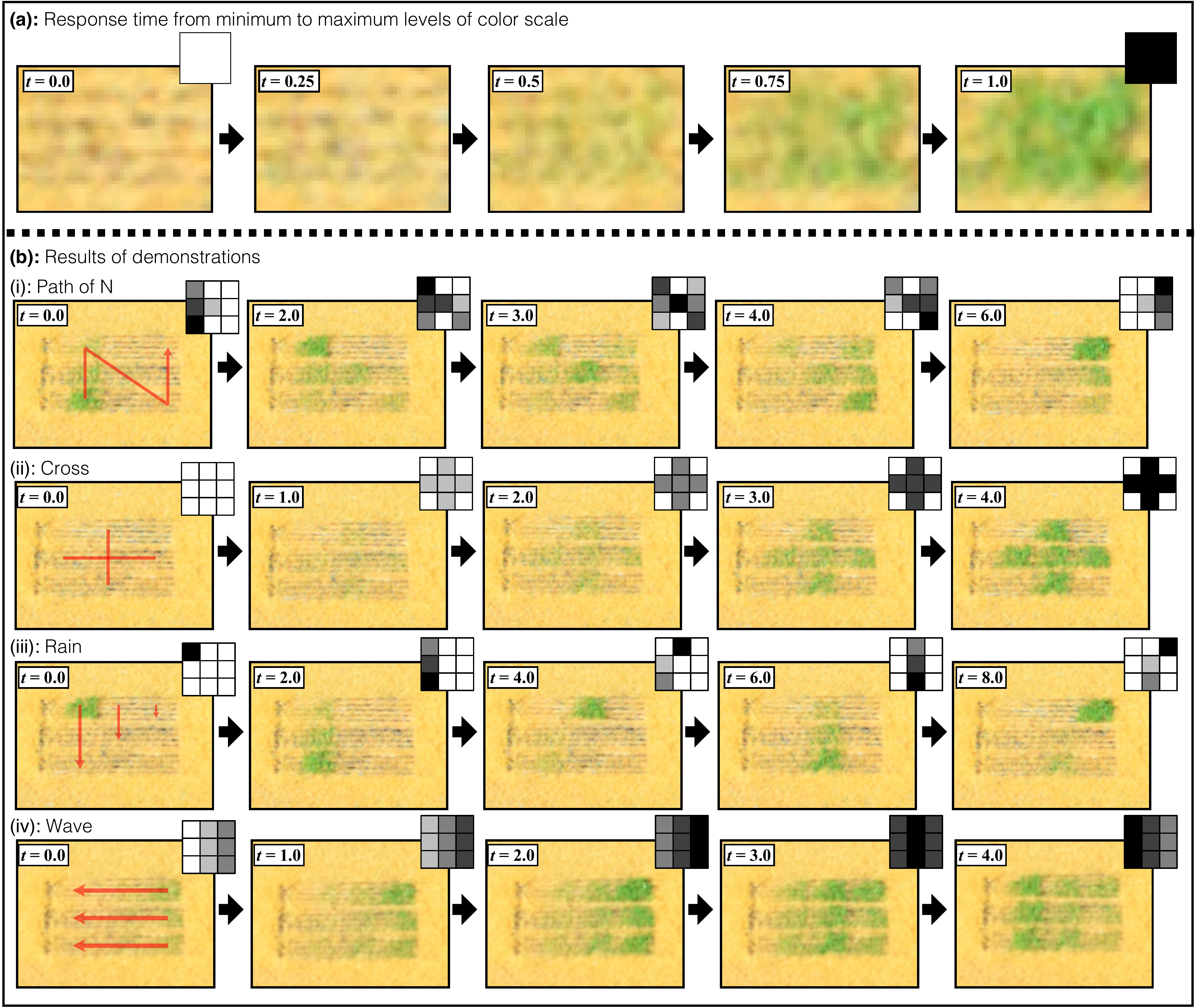}
  \caption{Demonstration results of a $3\times3$ pixels grass animation display using results of grass color scale setting experiments at $(h, d, \theta)=(1.7, 2.0, 0)$. (a) The response time to change the grass pixel from minimum and maximum of the grass color scale was 1.0 [s]. (b) Four animations of (i) path of letter N, (ii) cross, (iii) rain, and (iv) wave were played on the pixels grass animation display. This figure was created using Serif Affinity Designer version 1.10.0 (\url{https://affinity.serif.com/en-us/})}
  \label{demo}
\end{figure}

Figure \ref{demo}(a) shows the response time of changing the grass pixel from minimum to maximum of the color scale. The response time was 1.0 [s] that was within the keyframe interval of the animation system. Figure \ref{demo}(b) shows four demonstrations for the grass animation display. In the right upper corner of each grass animation display image, the corresponding input grayscale image is displayed. Figure \ref{demo}(b-i) shows the animation of the path of the letter N. The grass animation display plays the path animation as if the letter N is being written. Figure \ref{demo}(b-ii) shows an animation of a cross. The grayscale cross image turned gradually greener with time. Figure \ref{demo}(b-iii) shows an animation of the rainfall. In the demonstration, the raindrops fell from the left side of the grass animation display. Figure \ref{demo}(b-iv) shows the animation of a wave, which flowed from right to left of the grass animation display. Therefore, as shown in Figure \ref{demo}, the grass animation display can play several animations with the color scale using the results of the grass color scale setting experiments.

\section*{Limitation and future work}

\vspace{-5pt}

In this paper, the grass color scale can be controlled by adjusting the grass length of the grass pixel. On the other hand, another method of showing a color scale is spatial dithering. Using the spatial dithering, a wide range of the color scale can be expressed with a limited number of the color scale. In particular, it is possible to display intermediate colors even using the color scale with 2 levels like electronic ink (E ink). Although the spatial dithering using the color scale of 2 levels can help the grass animation display show the color scale without adjusting the grass length, multiple grass pixels are needed to show an input level of a single pixel. In this paper, it was difficult to use the spatial dithering in the grass animation display because the grass animation display had only nine grass pixels. However, with the spatial dithering, the more levels of the color scale the grass animation display can control, the wider the apparent color scale can be shown. In future work, we are going to develop a grass system with multiple motors to show a wide color scale using the spatial dithering and adjusting the grass length.

This paper showed the grass color characteristics of the grass pixel through the grass color scale experiments using the digital camera. Moreover, BRDF (Bidirectional Reflectance Distribution Function) model is also useful to confirm the grass color behavior of the grass pixel depending on the viewer's position. However, since the brightness and the angle of incidence of sunlight were not measublack in the grass color scale experiments, it was impossible to build the BRDF model of the grass pixel using the 36 camera data of the experiments. In the future, we are planning to measure and model the BRDF of the grass pixel based on the viewer's positions, the grass colors, and the grass length. 

In the demonstrations, the response time to change the grass pixel from minimum and maximum of the grass color scale was 1.0 [s]. However, a faster response time is needed to play the animations on the grass animation display smoothly. This problem can be solved using a high-speed motor with a rotary encoder system. In future work, we are going to develop a high-speed grass pixel system to play smooth animations. 

The grass color scale experiment results of \color{black}$\theta=60^\circ$ and \color{black} $90^\circ$ showed that the slits of the grass pixel affected the additive color mixing of the grass pixel. To solve this problem, the grass pixel needs to be such that the appearance of the gaps for moving the green grass length does change hardly when it is viewed at $\theta=0^\circ, 30^\circ, 60^\circ$, and $90^\circ$. For example, holes of matrix-like grids are similar in appearance when they are viewed from multiple positions. When the grass pixel was developed, we tried a grass pixel with holes of matrix-like grids. However, it was difficult to insert the green grass evenly into all of the holes of the grids. Thus, in this paper, the slits were adopted to facilitate the insertion of the green grass into the holes. In future work, we will design a grass pixel with the holes of the grids to widen the viewing angle of the grass color display. 

The color scale with five levels can be mapped to the green grass length through the grass color scale setting procedure. However, it is not evident whether the optimized grass pixel based on a camera position can show the grass colors corresponding to the color scale to a viewer at another position. Hence, to achieve a wide-field grass animation display, we must develop a grass color scale setting procedure to map the grass length to the color scale at multiple camera positions simultaneously. 

\color{black}

In addition, we conducted the grass pixel experiments under natural sunlight from 10:00 a.m. to 2:00 p.m. to evaluate the grass color under standard conditions. However, we did not reveal whether the experimental results of the color scale can be used except from 10:00 a.m. to 2:00 p.m., e.g., morning and evening environments. In the future, we will clarify a color scale mapping method of the grass pixel from morning to evening.

\color{black}

Furthermore, user evaluations are likewise important because the spatial additive mixing of the grass pixel is based on human perception. \color{black} We are planning user evaluations to reveal whether the grass pixel can display the color scale clearly based on psychophysical aspects. \color{black} Moreover, herein, we constructed the $3\times3$ pixels grass animation display, which can play several primitive animations with the color scale. In the future, we plan to create a grass animation display to show graphical animations such as texts and icons.

\vspace{-5pt}

\section*{Conclusion}

\vspace{-5pt}

We proposed a dynamic grass color scale display technique based on a grass length for a green landscape-friendly animation display. We designed a grass color scale setting procedure using image processing to map a color scale with five levels to the green grass length of the grass pixel. The grass color scale setting procedure yields the green grass length corresponding to the color scale using the measublack grass color values of the grass pixel at regular green grass length intervals. We experimented with the grass color scale setting procedure \color{black} under natural sunlight between 10:00 a.m. and 2:00 p.m. \color{black} at multiple camera positions based on a viewer's height, a distance, and a horizontal angle between the viewer and a grass pixel. As a result, we revealed that the grass pixel can be regarded to show the grass colors corresponding to the color scale at \color{black} 30 \color{black} camera positions out of 36, which can be mapped to the green grass length. \color{black} In addition, we also conducted the experiments on the repeatability and reliability of the grass pixel under natural sunlight between 10:00 a.m. and 2:00 p.m. at several camera positions. As a result, the experiments revealed the grass pixel can show the same grass color to several camera positions when the same grass length was repeatedly inputted. \color{black} Several demonstrations were constructed using the experimental results to play the animation with the color scale on the grass animation display. In the demonstrations, a $3\times3$ pixels grass animation display was built using nine grass pixels and played several animations (e.g., path of letter N, cross, rain, and wave). In our future study, we will aim to improve the grass color scale display technique based on human perception through user evaluations. Furthermore, we will develop a wide-field grass animation display to show graphical animations with the color scale, such as text and icons.

\section*{Methods}

\subsection*{Grass pixel implementation}

The hardware of the grass pixel was designed by the computer-aided design software (Autodesk FUSION 360). The limit switch plate, the guide plate, and the part of the surface plate were made of black acrylic sheets (thickness: 5 [mm]). The acrylic sheets were cut by the diode laser cutter (NEJE Master 2S plus). The laser cutter was operated by the laser cutter software (LASERGRBL). The 3D printed pin and the part of the surface plate were made by the fused deposition modeling (FDM) 3D printer (Creality Ender-3 Pro) using black filament (ANYCUBIC Filament, material: polylactic acid (PLA), diameter: $1.75\pm0.02$ [mm]). The 3D printer was operated by the 3D printer slicer software (Ultimaker Cura). The frameworks of the grass pixel system were made of aluminum frames (YUKI LABORATORY LECOFRAME). The printed circuit board (PCB) for the grass pixel was designed by the PCB design software (Autodesk Eagle).

\subsection*{Software of grass color scale setting procedure for grass color scale experiments}

The software of the grass color scale setting procedure was developed based in Python3. The digital camera was controlled by the camera management software (gPhoto2). The raw image was developed in sRGB color space using the Python raw image management software (rawpy). The RGB values of the developed image were corrected with the color correction Python library (PlantCV). The computer vision library (OpenCV) was used to crop the developed image and calculate the average CIELAB values of the evaluation image. \color{black} The image processing library (scikit-image) was used to calculate the color difference of CIEDE2000. The formula manipulation system (Mathematica) was used to calculate $\tfrac{t}{dt}\Delta E^*_{00} (G_0, P(t))$. \color{black}

\section*{Data availability}

All results from the grass color scale experiments are available from the corresponding author upon reasonable request.

\small

\begin{thebibliography}{10}
\urlstyle{rm}
\expandafter\ifx\csname url\endcsname\relax
  \def\url#1{\texttt{#1}}\fi
\expandafter\ifx\csname urlprefix\endcsname\relax\def\urlprefix{URL }\fi
\expandafter\ifx\csname doiprefix\endcsname\relax\def\doiprefix{DOI: }\fi
\providecommand{\bibinfo}[2]{#2}
\providecommand{\eprint}[2][]{\url{#2}}

\bibitem{SILVA2018697}
\bibinfo{author}{Silva, B.~N.}, \bibinfo{author}{Khan, M.} \&
  \bibinfo{author}{Han, K.}
\newblock \bibinfo{journal}{\bibinfo{title}{Towards sustainable smart cities: A
  review of trends, architectures, components, and open challenges in smart
  cities}}.
\newblock {\emph{\JournalTitle{Sustainable Cities and Society}}}
  \textbf{\bibinfo{volume}{38}}, \bibinfo{pages}{697--713},
  \doiprefix\url{https://doi.org/10.1016/j.scs.2018.01.053}
  (\bibinfo{year}{2018}).

\bibitem{surveyPervasive}
\bibinfo{author}{Clinch, S.}, \bibinfo{author}{Alexander, J.} \&
  \bibinfo{author}{Gehring, S.}
\newblock \bibinfo{journal}{\bibinfo{title}{A survey of pervasive displays for
  information presentation}}.
\newblock {\emph{\JournalTitle{IEEE Pervasive Computing}}}
  \textbf{\bibinfo{volume}{15}}, \bibinfo{pages}{14--22},
  \doiprefix\url{10.1109/MPRV.2016.55} (\bibinfo{year}{2016}).

\bibitem{Strategies}
\bibinfo{author}{Hespanhol, L.} \& \bibinfo{author}{Tomitsch, M.}
\newblock \bibinfo{journal}{\bibinfo{title}{{Strategies for Intuitive
  Interaction in Public Urban Spaces}}}.
\newblock {\emph{\JournalTitle{Interacting with Computers}}}
  \textbf{\bibinfo{volume}{27}}, \bibinfo{pages}{311--326},
  \doiprefix\url{10.1093/iwc/iwu051} (\bibinfo{year}{2015}).
\newblock
  \eprint{https://academic.oup.com/iwc/article-pdf/27/3/311/6957736/iwu051.pdf}.

\bibitem{alt2016Opportunistic}
\bibinfo{author}{Alt, F.} \& \bibinfo{author}{Vehns, J.}
\newblock \bibinfo{title}{Opportunistic deployments: Challenges and
  opportunities of conducting public display research at an airport}.
\newblock In \emph{\bibinfo{booktitle}{Proceedings of the 5th {{ACM
  International Symposium}} on {{Pervasive Displays}}}}, {{PerDis}} '16,
  \bibinfo{pages}{106--117}, \doiprefix\url{10.1145/2914920.2915020}
  (\bibinfo{publisher}{{Association for Computing Machinery}},
  \bibinfo{address}{{New York, NY, USA}}, \bibinfo{year}{2016}).

\bibitem{coenen2016Synchronized}
\bibinfo{author}{Coenen, J.}, \bibinfo{author}{Wouters, N.} \&
  \bibinfo{author}{Moere, A.~V.}
\newblock \bibinfo{title}{Synchronized wayfinding on multiple consecutively
  situated public displays}.
\newblock In \emph{\bibinfo{booktitle}{Proceedings of the 5th {{ACM
  International Symposium}} on {{Pervasive Displays}}}}, {{PerDis}} '16,
  \bibinfo{pages}{182--196}, \doiprefix\url{10.1145/2914920.2929906}
  (\bibinfo{publisher}{{Association for Computing Machinery}},
  \bibinfo{address}{{New York, NY, USA}}, \bibinfo{year}{2016}).

\bibitem{yoshino2020KI}
\bibinfo{author}{Yoshino, K.} \emph{et~al.}
\newblock \bibinfo{title}{{{KI}}/{{OSK}}: {{Practice Study}} of {{Load
  Sensitive Board}} for {{Farmers Market}}}.
\newblock In \emph{\bibinfo{booktitle}{Extended {{Abstracts}} of the 2020 {{CHI
  Conference}} on {{Human Factors}} in {{Computing Systems}}}}, {{CHI EA}} '20,
  \bibinfo{pages}{1--8}, \doiprefix\url{10.1145/3334480.3375208}
  (\bibinfo{publisher}{{Association for Computing Machinery}},
  \bibinfo{address}{{New York, NY, USA}}, \bibinfo{year}{2020}).

\bibitem{muta2015Interactive}
\bibinfo{author}{Muta, M.}, \bibinfo{author}{Masuko, S.},
  \bibinfo{author}{Shinzato, K.} \& \bibinfo{author}{Mujibiya, A.}
\newblock \bibinfo{title}{Interactive {{Study}} of {{WallSHOP}}: {{Multiuser
  Connectivity}} between {{Public Digital Advertising}} and {{Private Devices}}
  for {{Personalized Shopping}}}.
\newblock In \emph{\bibinfo{booktitle}{Proceedings of the 4th {{International
  Symposium}} on {{Pervasive Displays}}}}, {{PerDis}} '15,
  \bibinfo{pages}{187--193}, \doiprefix\url{10.1145/2757710.2757732}
  (\bibinfo{publisher}{{Association for Computing Machinery}},
  \bibinfo{address}{{New York, NY, USA}}, \bibinfo{year}{2015}).

\bibitem{yoo2020Putting}
\bibinfo{author}{Yoo, D.} \emph{et~al.}
\newblock \bibinfo{title}{Putting {{Down Roots}}: {{Exploring}} the
  {{Placeness}} of {{Virtual Collections}} in {{Public Libraries}}}.
\newblock In \emph{\bibinfo{booktitle}{Proceedings of the 2020 {{ACM Designing
  Interactive Systems Conference}}}}, \bibinfo{pages}{723--734}
  (\bibinfo{publisher}{{Association for Computing Machinery}},
  \bibinfo{address}{{New York, NY, USA}}, \bibinfo{year}{2020}).

\bibitem{kukka2018UbiLibrary}
\bibinfo{author}{Kukka, H.}, \bibinfo{author}{Heikkinen, T.},
  \bibinfo{author}{Kyt{\"o}kangas, H.}, \bibinfo{author}{Tanska, T.} \&
  \bibinfo{author}{Ojala, T.}
\newblock \bibinfo{title}{{{UbiLibrary}}: {{Situated Large Public Display}} as
  {{Interactive Interface}} to {{Library Services}}}.
\newblock In \emph{\bibinfo{booktitle}{Proceedings of the 22nd {{International
  Academic Mindtrek Conference}}}}, Mindtrek '18, \bibinfo{pages}{192--201},
  \doiprefix\url{10.1145/3275116.3275143} (\bibinfo{publisher}{{Association for
  Computing Machinery}}, \bibinfo{address}{{New York, NY, USA}},
  \bibinfo{year}{2018}).

\bibitem{claes2016Bicycle}
\bibinfo{author}{Claes, S.}, \bibinfo{author}{Slegers, K.} \&
  \bibinfo{author}{Vande~Moere, A.}
\newblock \bibinfo{title}{The {{Bicycle Barometer}}: {{Design}} and
  {{Evaluation}} of {{Cyclist-Specific Interaction}} for a {{Public Display}}}.
\newblock In \emph{\bibinfo{booktitle}{Proceedings of the 2016 {{CHI
  Conference}} on {{Human Factors}} in {{Computing Systems}}}}, {{CHI}} '16,
  \bibinfo{pages}{5824--5835}, \doiprefix\url{10.1145/2858036.2858429}
  (\bibinfo{publisher}{{Association for Computing Machinery}},
  \bibinfo{address}{{New York, NY, USA}}, \bibinfo{year}{2016}).

\bibitem{coenen2021Public}
\bibinfo{author}{Coenen, J.} \& \bibinfo{author}{Moere, A.~V.}
\newblock \bibinfo{journal}{\bibinfo{title}{Public {{Data Visualization}}:
  {{Analyzing Local Running Statistics}} on {{Situated Displays}}}}.
\newblock {\emph{\JournalTitle{Computer Graphics Forum}}}
  \textbf{\bibinfo{volume}{40}}, \bibinfo{pages}{159--171},
  \doiprefix\url{10.1111/cgf.14297} (\bibinfo{year}{2021}).

\bibitem{chang2014Lunch}
\bibinfo{author}{Chang, K. S.-P.}, \bibinfo{author}{Danis, C.~M.} \&
  \bibinfo{author}{Farrell, R.~G.}
\newblock \bibinfo{title}{Lunch line: Using public displays and mobile devices
  to encourage healthy eating in an organization}.
\newblock In \emph{\bibinfo{booktitle}{Proceedings of the 2014 {{ACM
  International Joint Conference}} on {{Pervasive}} and {{Ubiquitous
  Computing}}}}, {{UbiComp}} '14, \bibinfo{pages}{823--834},
  \doiprefix\url{10.1145/2632048.2636086} (\bibinfo{publisher}{{Association for
  Computing Machinery}}, \bibinfo{address}{{New York, NY, USA}},
  \bibinfo{year}{2014}).

\bibitem{altmeyer2018Extending}
\bibinfo{author}{Altmeyer, M.}, \bibinfo{author}{Lessel, P.},
  \bibinfo{author}{Sander, T.} \& \bibinfo{author}{Kr{\"u}ger, A.}
\newblock \bibinfo{title}{Extending a {{Gamified Mobile App}} with a {{Public
  Display}} to {{Encourage Walking}}}.
\newblock In \emph{\bibinfo{booktitle}{Proceedings of the 22nd {{International
  Academic Mindtrek Conference}}}}, Mindtrek '18, \bibinfo{pages}{20--29},
  \doiprefix\url{10.1145/3275116.3275135} (\bibinfo{publisher}{{Association for
  Computing Machinery}}, \bibinfo{address}{{New York, NY, USA}},
  \bibinfo{year}{2018}).

\bibitem{FreeStanding}
\bibinfo{author}{{THE RUGGED DISPLAY COMPANY}}.
\newblock \bibinfo{title}{Free-standing digital signage podium monitors at
  manchester science park's new bright building}.
\newblock
  \bibinfo{howpublished}{\url{https://www.flatvision.co.uk/portfolio-item/podium-monitors-msp-case-study/}}.
\newblock \bibinfo{note}{Accessed: 2022-03-01}.

\bibitem{northerncrossSosei}
\bibinfo{author}{{Hokkaido Magazine KAI}}.
\newblock \bibinfo{title}{Sosei {{River Park Symphony}}}.
\newblock
  \bibinfo{howpublished}{\url{http://kai-hokkaido.com/en/feature_vol34_ichiba8/}}.
\newblock \bibinfo{note}{Accessed: 2022-03-01}.

\bibitem{maller2006Healthy}
\bibinfo{author}{Maller, C.}, \bibinfo{author}{Townsend, M.},
  \bibinfo{author}{Pryor, A.}, \bibinfo{author}{Brown, P.} \&
  \bibinfo{author}{St~Leger, L.}
\newblock \bibinfo{journal}{\bibinfo{title}{Healthy nature healthy people:
  `contact with nature' as an upstream health promotion intervention for
  populations}}.
\newblock {\emph{\JournalTitle{Health Promotion International}}}
  \textbf{\bibinfo{volume}{21}}, \bibinfo{pages}{45--54},
  \doiprefix\url{10.1093/heapro/dai032} (\bibinfo{year}{2006}).

\bibitem{abraham2010Landscape}
\bibinfo{author}{Abraham, A.}, \bibinfo{author}{Sommerhalder, K.} \&
  \bibinfo{author}{Abel, T.}
\newblock \bibinfo{journal}{\bibinfo{title}{Landscape and well-being: A scoping
  study on the health-promoting impact of outdoor environments}}.
\newblock {\emph{\JournalTitle{International Journal of Public Health}}}
  \textbf{\bibinfo{volume}{55}}, \bibinfo{pages}{59--69},
  \doiprefix\url{10.1007/s00038-009-0069-z} (\bibinfo{year}{2010}).

\bibitem{lee2017Experimental}
\bibinfo{author}{Lee, J.}
\newblock \bibinfo{journal}{\bibinfo{title}{Experimental {{Study}} on the
  {{Health Benefits}} of {{Garden Landscape}}}}.
\newblock {\emph{\JournalTitle{International Journal of Environmental Research
  and Public Health}}} \textbf{\bibinfo{volume}{14}},
  \doiprefix\url{10.3390/ijerph14070829} (\bibinfo{year}{2017}).

\bibitem{robinson2019Sustainabot}
\bibinfo{author}{Robinson, S.}, \bibinfo{author}{Pearson, J.},
  \bibinfo{author}{Holton, M.~D.}, \bibinfo{author}{Ahire, S.} \&
  \bibinfo{author}{Jones, M.}
\newblock \bibinfo{title}{Sustainabot - {{Exploring}} the {{Use}} of {{Everyday
  Foodstuffs}} as {{Output}} and {{Input}} for and with {{Emergent Users}}}.
\newblock In \emph{\bibinfo{booktitle}{Proceedings of the 2019 {{CHI
  Conference}} on {{Human Factors}} in {{Computing Systems}}}},
  \bibinfo{pages}{1--12} (\bibinfo{publisher}{{Association for Computing
  Machinery}}, \bibinfo{address}{{New York, NY, USA}}, \bibinfo{year}{2019}).

\bibitem{nagafuchi2020Polka}
\bibinfo{author}{Nagafuchi, R.} \emph{et~al.}
\newblock \bibinfo{title}{Polka: {{A Water-jet Printer}} for {{Painting}} on
  the {{Grounds}}}.
\newblock In \emph{\bibinfo{booktitle}{Proceedings of the {{International
  Conference}} on {{Advanced Visual Interfaces}}}}, \bibinfo{number}{63},
  \bibinfo{pages}{1--5} (\bibinfo{publisher}{{Association for Computing
  Machinery}}, \bibinfo{address}{{New York, NY, USA}}, \bibinfo{year}{2020}).

\bibitem{sugiura2014Graffiti}
\bibinfo{author}{Sugiura, Y.} \emph{et~al.}
\newblock \bibinfo{title}{Graffiti fur: Turning your carpet into a computer
  display}.
\newblock In \emph{\bibinfo{booktitle}{Proceedings of the 27th Annual {{ACM}}
  Symposium on {{User}} Interface Software and Technology}}, {{UIST}} '14,
  \bibinfo{pages}{149--156}, \doiprefix\url{10.1145/2642918.2647370}
  (\bibinfo{publisher}{{Association for Computing Machinery}},
  \bibinfo{address}{{New York, NY, USA}}, \bibinfo{year}{2014}).

\bibitem{2015Rice}
\bibinfo{author}{{aptinet Aomori Sightseeing Guide}}.
\newblock \bibinfo{title}{Rice {{Paddy Art}} -{{Inakadate Village}}}.
\newblock
  \bibinfo{howpublished}{\url{https://www.en-aomori.com/scenery-025.html}}.
\newblock \bibinfo{note}{Accessed: 2022-03-01}.

\bibitem{sugiura2017Grassffiti}
\bibinfo{author}{Sugiura, Y.} \emph{et~al.}
\newblock \bibinfo{title}{Grassffiti: {{Drawing Method}} to {{Produce
  Large-scale Pictures}} on {{Conventional Grass Fields}}}.
\newblock In \emph{\bibinfo{booktitle}{Proceedings of the {{Eleventh
  International Conference}} on {{Tangible}}, {{Embedded}}, and {{Embodied
  Interaction}}}}, {{TEI}} '17, \bibinfo{pages}{413--417},
  \doiprefix\url{10.1145/3024969.3025067} (\bibinfo{publisher}{{Association for
  Computing Machinery}}, \bibinfo{address}{{New York, NY, USA}},
  \bibinfo{year}{2017}).

\bibitem{AirPrint}
\bibinfo{author}{{New Ground Technology}}.
\newblock \bibinfo{title}{{{AirPrint}}, gently print logos on turf, pitch or
  grass using only air. {{Big}} impact, {{Erasable}} in minutes}.
\newblock
  \bibinfo{howpublished}{\url{https://www.newgroundtechnology.com/airprint}}.

\bibitem{scheible2016DroneLandArt}
\bibinfo{author}{Scheible, J.} \& \bibinfo{author}{Funk, M.}
\newblock \bibinfo{title}{{{DroneLandArt}}: Landscape as organic pervasive
  display}.
\newblock In \emph{\bibinfo{booktitle}{Proceedings of the 5th {{ACM
  International Symposium}} on {{Pervasive Displays}}}}, {{PerDis}} '16,
  \bibinfo{pages}{255--256}, \doiprefix\url{10.1145/2914920.2939883}
  (\bibinfo{publisher}{{Association for Computing Machinery}},
  \bibinfo{address}{{New York, NY, USA}}, \bibinfo{year}{2016}).

\bibitem{Big}
\bibinfo{author}{{Ackroyd \& Harvey}}.
\newblock \bibinfo{title}{Big chill}.
\newblock
  \bibinfo{howpublished}{\url{https://www.ackroydandharvey.com/big-chill/}}.
\newblock \bibinfo{note}{Accessed: 2022-03-01}.

\bibitem{minuto2012Growing}
\bibinfo{author}{Minuto, A.} \& \bibinfo{author}{Nijholt, A.}
\newblock \bibinfo{title}{Growing grass: A smart material interactive display,
  design and construction history}.
\newblock In \emph{\bibinfo{booktitle}{Proceedings of the 1st Workshop on
  {{Smart Material Interfaces}}: {{A Material Step}} to the {{Future}}}},
  {{SMI}} '12, \bibinfo{pages}{1--5}, \doiprefix\url{10.1145/2459056.2459063}
  (\bibinfo{publisher}{{Association for Computing Machinery}},
  \bibinfo{address}{{New York, NY, USA}}, \bibinfo{year}{2012}).

\bibitem{ohkubo2013Interface}
\bibinfo{author}{Ohkubo, M.}, \bibinfo{author}{Ooide, Y.} \&
  \bibinfo{author}{Nojima, T.}
\newblock \bibinfo{title}{An interface composed of a collection of "smart
  hairs"}.
\newblock In \emph{\bibinfo{booktitle}{Proceedings of the Second International
  Workshop on {{Smart}} Material Interfaces: Another Step to a Material
  Future}}, {{SMI}} '13, \bibinfo{pages}{23--26},
  \doiprefix\url{10.1145/2534688.2534692} (\bibinfo{publisher}{{Association for
  Computing Machinery}}, \bibinfo{address}{{New York, NY, USA}},
  \bibinfo{year}{2013}).

\bibitem{tanaka2021Natural}
\bibinfo{author}{Tanaka, K.}, \bibinfo{author}{Mikawa, M.} \&
  \bibinfo{author}{Fujisawa, M.}
\newblock \bibinfo{title}{Natural {{Landscape-Friendly Animation Display
  Technique Using Shape-Changing Artificial Grass System}}}.
\newblock In \emph{\bibinfo{booktitle}{2021 {{IEEE International Conference}}
  on {{Systems}}, {{Man}}, and {{Cybernetics}} ({{SMC}})}},
  \bibinfo{pages}{2549--2554}, \doiprefix\url{10.1109/SMC52423.2021.9659032}
  (\bibinfo{year}{2021}).

\bibitem{sharma2005ciede2000}
\bibinfo{author}{Sharma, G.}, \bibinfo{author}{Wu, W.} \&
  \bibinfo{author}{Dalal, E.~N.}
\newblock \bibinfo{journal}{\bibinfo{title}{The ciede2000 color-difference
  formula: Implementation notes, supplementary test data, and mathematical
  observations}}.
\newblock {\emph{\JournalTitle{Color Research \& Application: Endorsed by
  Inter-Society Color Council, The Colour Group (Great Britain), Canadian
  Society for Color, Color Science Association of Japan, Dutch Society for the
  Study of Color, The Swedish Colour Centre Foundation, Colour Society of
  Australia, Centre Fran{\c{c}}ais de la Couleur}}}
  \textbf{\bibinfo{volume}{30}}, \bibinfo{pages}{21--30}
  (\bibinfo{year}{2005}).

\bibitem{sato1997Requirements}
\bibinfo{author}{Sato, T.} \& \bibinfo{author}{Sato, I.}
\newblock \bibinfo{title}{Requirements {{Governing Light Sources Used}} for
  {{Color Evaluation}}}.
\newblock In \emph{\bibinfo{booktitle}{Proceedings of the 8th {{Congress}} of
  the {{International Colour Association}}}}, vol.~\bibinfo{volume}{1} of
  \emph{\bibinfo{series}{{{AIC Color}} '97}}, \bibinfo{pages}{415--418}
  (\bibinfo{year}{1997}).

  \bibitem{an2012graphic}
  \bibinfo{author}{ANSI/Committee for Graphic Arts Technologies Standards (CGATS).}
  \newblock \bibinfo{journal}{\bibinfo{title}{CGATS TR 016-2014 Graphic technology—printing
    tolerance and conformity assessment}}.
  \newblock (\bibinfo{year}{2014}).


\end{thebibliography}


\normalsize

\section*{Acknowledgments}
This work was supported by JST SPRING, Grant Number JTMJST2124.

\section*{Author contributions}

K.T. designed and implemented the grass color control system. K.T., Y.K., \color{black} A.M., \color{black} M.M., and M.F. conceived the grass gradation scale setting procedure. Y.K. \color{black} and A.M. \color{black} implemented the software of the grass color scale setting procedure. K.T., Y.K. \color{black} and A.M. \color{black} conducted the outdoor experiments of the grass color scale setting procedure. K.T. conducted the animation demonstrations. All authors discussed the results. K.T. wrote the initial manuscript, and M.M reviewed the manuscript.

\section*{Competing interests}
The authors declare no competing interests.

\end{document}